\documentclass[useAMS,usenatbib,a4paper]{mn2e}
\usepackage{graphicx}
\usepackage{amsmath}
\usepackage{txfonts}

\def\nodata{\dots}
\def\aap{A\&A}
\def\apjl{ApJ}
\def\aph{APh}
\def\apjs{ApJS}
\def\apj{ApJ}
\def\araa{ARA\&A}
\def\nat{Nat}
\def\sci{Sci}
\def\baas{BAAS}
\def\mnras{MNRAS}
\def\prl{PRL}

\title[Synoptic VHE blazar studies]{Synoptic studies of seventeen blazars detected in very high-energy gamma-rays}
\author[R. M. Wagner]{R. M. Wagner\thanks{E-mail: robert.wagner@mppmu.mpg.de}
\\
Max-Planck-Institut f\"ur Physik, F\"ohringer Ring 6, D-80805
M\"unchen, Germany}

\date{Accepted 2007 December 12. Received 2007 December 10; in original form 2007 September 13}

\pubyear{2008}

\begin{document}
\maketitle
\label{firstpage}

\begin{abstract}
Since 2002, the number of detected blazars at gamma-ray energies
above 100 GeV has more than doubled. I study 17 blazars currently
known to emit $E>100$~GeV gamma rays. Their intrinsic energy
spectra are reconstructed by removing extragalactic background
light attenuation effects. Luminosity and spectral slope in the
$E>100$~GeV region are then compared and correlated among each
other, with X-ray, optical and radio data, and with the estimated
black hole (BH) masses of the respective host galaxies.

According to expectations from synchrotron self-Compton emission
models, a correlation on the 3.6-$\sigma$ significance level between gamma-ray
and X-ray fluxes is found, while correlations between gamma-ray
and optical/radio fluxes are less pronounced. Further, a general
hardening of the blazar spectra in the $E>100$~GeV region with
increasing gamma-ray luminosity is observed, both for the full
17-source sample and for those sources which have been detected at
distinct flux levels. This goes in line with a correlation of the
gamma-ray luminosity and the synchrotron peak frequency, which is
also seen. Tests for possible selection effects reveal a hardening
of the spectra with increasing redshift. The blazar gamma-ray
emission might depend on the mass of the central BH. The blazars
under study do, however, show no correlation of the BH masses with
the spectral index and the luminosity in the $E>100$~GeV region.

I also consider temporal properties of the X-ray and $E>100$~GeV
gamma-ray flux. No general trends are found, except for the
observation that the blazars with the most massive BHs do not show
particularly high duty cycles. These blazars include Mkn\,501 and
PKS\,2155-304, for which recently very fast flares have been
reported. In general, VHE flare time-scales are not found to scale
with the BH mass.

As a specific application of the luminosity study, a constraint
for the still undetermined redshift of the blazar PG\,1553+113 is
discussed.
\end{abstract}


\section{Introduction}
All but one of the detected extragalactic very high energy (VHE,
defined by $E>100\,\mathrm{GeV}$) gamma ($\gamma$) ray sources so
far are blazars. Within the unified scheme
\citep[e.g.][]{UrryPadovani} of active galactic nuclei (AGN),
blazars comprise the rare and extreme subclasses of BL~Lac objects
and flat spectrum radio quasars (FSRQs). These are characterised
by high apparent luminosities, short variability time-scales, and
apparent superluminal motion of jet components. These observations
can be explained by highly relativistic, beamed plasma outflows
(jets) closely aligned to the observer's line of sight
\citep{1979ApJ...232...34B} powered by central supermassive black
holes accreting at sub-Eddington rates
\citep{1969Natur.223..690L,1978Natur.275..516R}. The prime
scientific interest in VHE $\gamma$-ray emitting blazars (in the
following, `VHE blazars') is twofold: (1) To understand the
particle acceleration and $\gamma$-ray production mechanisms,
assumed to take place in the jets and to be linked to the central
supermassive black hole (BH). Knowledge of the VHE
emission process will also contribute to the further understanding
of the accretion processes in AGN, jet formation processes, and
the jet structure. (2) To use the VHE $\gamma$-rays as a probe of
the extragalactic background light \citep[EBL;
e.g.][]{hauser,2005PhR...409..361K} spectrum in the wavelength
range between about $0.3$ to $30~\mu\rm m$. Determining the EBL
spectrum in this wavelength range may allow to constrain the star
formation rate (convolved with the initial mass function) in the
early Universe. In order to assess both issues, it is essential to
have a large sample of VHE $\gamma$-ray blazars at hand. Ideally
it should encompass a wide range in redshift for EBL studies and
at the same time include groups of sources at similar distances in
order to probe and compare properties of the individual sources
without possible systematic uncertainties caused by the EBL
de-absorption.
\begin{figure*}
\center{\includegraphics[width=0.65\linewidth]{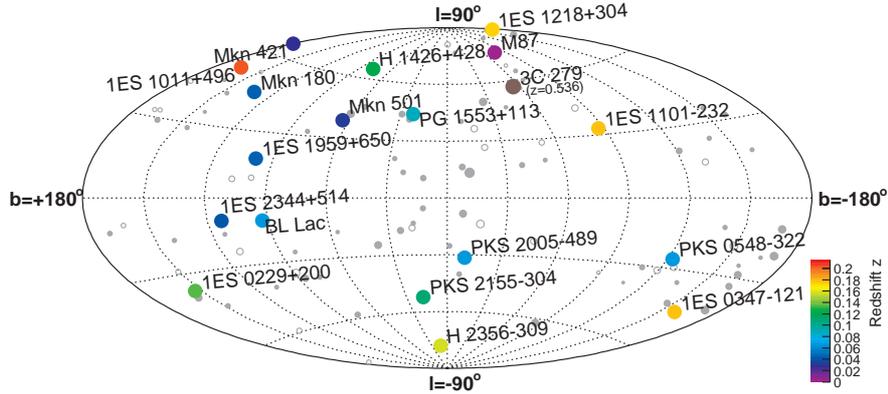}
\caption{Currently known VHE $\gamma$-ray blazars along with the
identified (66 objects) and tentatively identified AGN (27
objects) in the 3rd EGRET catalogue of $\gamma$-ray sources (solid
grey dots: identified AGN; open grey dots: tentatively identified
AGN). EGRET data from \citet{hartman}. 
The sources are shown in a galactic coordinate
system.\label{fig:comp:sp}}}
\end{figure*}

The preconditions for such studies have much improved recently:
Before 2004, only a few nearby extragalactic sources had been
established as VHE $\gamma$-ray emitters \citep[e.g.][]{Moriicrc}
and provided hardly enough data to perform comparative studies.
Around 2004, the third generation of imaging air Cerenkov
telescopes (IACTs, e.g. MAGIC, \citealt{Baixeras,Cortina} and
H.E.S.S., \citealt{HESStech}), the most successful tools so far to
explore VHE $\gamma$-rays, started to deliver scientific results.
To date, the VHE blazar sample with available spectral information in the
VHE region comprises 17 BL Lac objects, 
among them one LBL object, BL~Lacertae. Furthermore, now the
redshifts of the known VHE blazars reach up to $z=0.212$ -- or even
to $z=0.536$, considering the recently announced discovery of the
first FSRQ in VHE $\gamma$-rays, 3C~279 \citep{279}. For
3C~279, however, no VHE spectrum has been published yet, therefore
the VHE luminosity and spectral slope of 3C~279 are not yet
available for this study. M87, a FR~I radio galaxy also detected
in VHE $\gamma$-rays, is not included in the study, as its VHE
$\gamma$-ray production mechanism may differ from that in blazars
\citep{hess87n}. Fig.~\ref{fig:comp:sp} shows the sky positions of
the known VHE blazars in galactic coordinates along with the AGN
identified in the 3rd EGRET catalogue.

The electromagnetic continuum spectra of blazars extend over many
orders of magnitude from radio frequencies to sometimes multi-TeV
energies and are dominated by non-thermal emission that consists
in a $\nu F_\nu$ representation of two pronounced peaks. The
low-energy peak, located between the IR and hard X-rays, is
thought to arise from synchrotron emission of ultrarelativistic
electrons, accelerated by shocks moving along the jets at
relativistic bulk speed. Depending on the location of the
low-energy peak, BL~Lac objects are often referred to as
high-frequency peaked (HBL; in the UV to X-ray domain) or
low-frequency peaked (LBL; in the near-IR to optical) BL~Lac
objects \citep{fossati}, although the transition is smooth rather
than dichotomic. The origin of the high-energy peak at MeV to TeV
energies is still debated. It is commonly explained by inverse
Compton upscattering of low-energy photons by electrons. The seed
photons may originate from synchrotron radiation produced by the
same electron population \citep*[synchrotron-self Compton (SSC)
models; e.g.][]{maraschi,coppi2} or belong to ambient thermal
photon fields \citep*[external inverse Compton models;
e.g.][]{mk89,1994ApJ...421..153S,1994ApJS...90..945D}. In hadronic
models, which can also explain the observed features, interactions
of a highly relativistic jet outflow with ambient matter
\citep{1997ApJ...478L...5D,1993ApJ...402L..29B}, proton-induced
cascades \citep{MannheimProton}, synchrotron radiation by protons
\citep{AharonianHad,MueckeProtheroe}, or curvature radiation, are
responsible for the high energy photons. The AGN identified in
the EGRET data are predominantly powerful FSRQs and quasars with
SEDs peaking at rather low frequencies, and thus only few of these
(Mkn 421, PKS 2155-304, BL~Lacertae and 3C~279) were also
detected in the VHE range.

Knowing the variability time-scales of the VHE $\gamma$-ray
emission and the form of the two-bump spectral energy distribution
(SED) enables the derivation of all input parameters of one-zone
SSC models \citep{Tavecchio1998}, which describe the observed
emission in BL~Lac objects reasonably well. Strictly simultaneous
and temporally-resolved measurements of the SED, however, are only
rarely possible and more often than not are also severely
restricted by the (temporal) instrumental resolution. Generally,
detailed spectral studies particularly in low-emission states are
rather demanding. In addition, the determination of the location
of the high-energy peak $\nu^\mathrm{IC}_\mathrm{peak}$ generally
would require complementary satellite detector coverage of the SED
between some hundred MeV and $\approx 50$~GeV. This region of the
SED, however, is difficult to access due to the low fluxes
expected (up to $\approx 100$ MeV) from extragalactic sources in
between the two bumps and due to insufficient instrumental
sensitivity for energies exceeding some GeVs.

In this paper, for the first time studies of the VHE emission
properties of the complete set of all currently known VHE blazars
(cf. Tab.~\ref{tab:TeVSources}) are performed.
First, the detected VHE blazars are brought into context with the
AGN searches conducted by IACTs so far and the expected
$\gamma$-ray attenuation by the EBL in Sect.~\ref{sect:grhorsect}.
After a study of the black hole mass distribution of the VHE
blazars in Sect.~\ref{sect:comp:bhmavg}, I infer intrinsic
emission properties in the VHE $\gamma$-ray regime in
Sect.~\ref{sect:intrvhe}. Because measurements of the location and
shape of the high-energy bump are elusive at present for almost
all VHE blazars, the observed $\gamma$-ray luminosity and spectral
slope in the VHE region are used as auxiliary observables to
characterise the VHE $\gamma$-ray emission. The main part of the
paper (Sect.~\ref{sect:corstudies1}--\ref{sect:intrinsicspectra}) is devoted to the search for
correlations of these observables with the X-ray emission
properties, the optical and the radio luminosity, and with the
black hole mass estimations. In Sect.~\ref{sect:comp:501lumilim}
the VHE blazar luminosity distribution is used to address the
specific problem of the unknown redshift of PG\,1553+113 by
deriving an upper redshift for this VHE blazar. Finally,
Sect.~\ref{sect:dcsect} turns to the study of X-ray and VHE
$\gamma$-ray timing properties. Sect.~\ref{sect:conc} summarises
the main conclusions of the studies.

\begin{table}
\caption{Extragalactic VHE $\gamma$-ray sources, listed in
chronological order of their discovery.} \label{tab:TeVSources}
\begin{tabular}{lccl}
\hline
Source & Type & Redshift $z$ & Discovery reference  \\
\hline
Mkn\,421       & HBL & 0.030\phantom{0} & \citet{Punch1992}  \\
Mkn\,501       & HBL & 0.034\phantom{0} & \citet{Quinn1996} \\
1ES\,2344+514  & HBL & 0.044\phantom{0} & \citet{Catanese1998}  \\
1ES\,1959+650  & HBL & 0.047\phantom{0} & \citet{Nishiyama1999}  \\
PKS\,2155-304  & HBL & 0.116\phantom{0} & \citet{Chadwick}  \\
H\,1426+428    & HBL & 0.129\phantom{0} & \citet{Horan}  \\
M87$^a$            & FR~I  & 0.0044           & \citet{Tluczykont}\\
\hline
PKS\,2005-489  & HBL &  0.071\phantom{0}& \citet{HESS2005}   \\
1ES\,1218+304 & HBL & 0.182\phantom{0}  & \citet{MAGIC1218} \\
H\,2356-309   & HBL & 0.165\phantom{0} & \citet{HESSAGNNature}   \\
1ES\,1101-232  & HBL & 0.186\phantom{0} & \citet{HESSAGNNature}  \\
PG\,1553+113   & HBL & $^b$            & \citet{HESS1553}, \\
               &     &               & \citet{MAGIC1553published}  \\
Mkn\,180       & HBL & 0.045\phantom{0} & \citet{MAGIC180}   \\
PKS\,0548-322  & HBL & 0.069\phantom{0} & \citet{0548icrc} \\
BL~Lacertae    & LBL & 0.069\phantom{0} & \citet{MAGICBLLac} \\
1ES\,1011+496  & HBL & 0.212\phantom{0} & \citet{1011} \\
1ES\,0229+200  & HBL & 0.139\phantom{0} & \citet{0229} \\
1ES\,0347-121  & HBL & 0.188\phantom{0} & \citet{0347} \\
3C~279$^c$     & FSRQ& 0.536\phantom{0} & \citet{279} \\
\hline
\end{tabular}

\medskip
The upper part of the table shows the confirmed sources prior to
the advent of new generation instruments like MAGIC and H.E.S.S.,
while the lower panel summarises the sources discovered after
2002. HBL: High-frequency peaked BL~Lac object, LBL: Low-frequency
peaked BL~Lac object, FSRQ: Flat spectrum radio quasar, FR:
Fanaroff--Riley galaxy. $^a$ M87 is not included in the present
study. $^b$ This redshift is currently under discussion, cf. Sect.
\ref{sect:comp:501lumilim}. $^c$ No spectrum of 3C~279 has been
published yet.
\end{table}

\section{Population studies and the $\gamma$-ray horizon}
\label{sect:grhorsect}

\begin{figure*}
\center{
\includegraphics[width=0.495\linewidth]{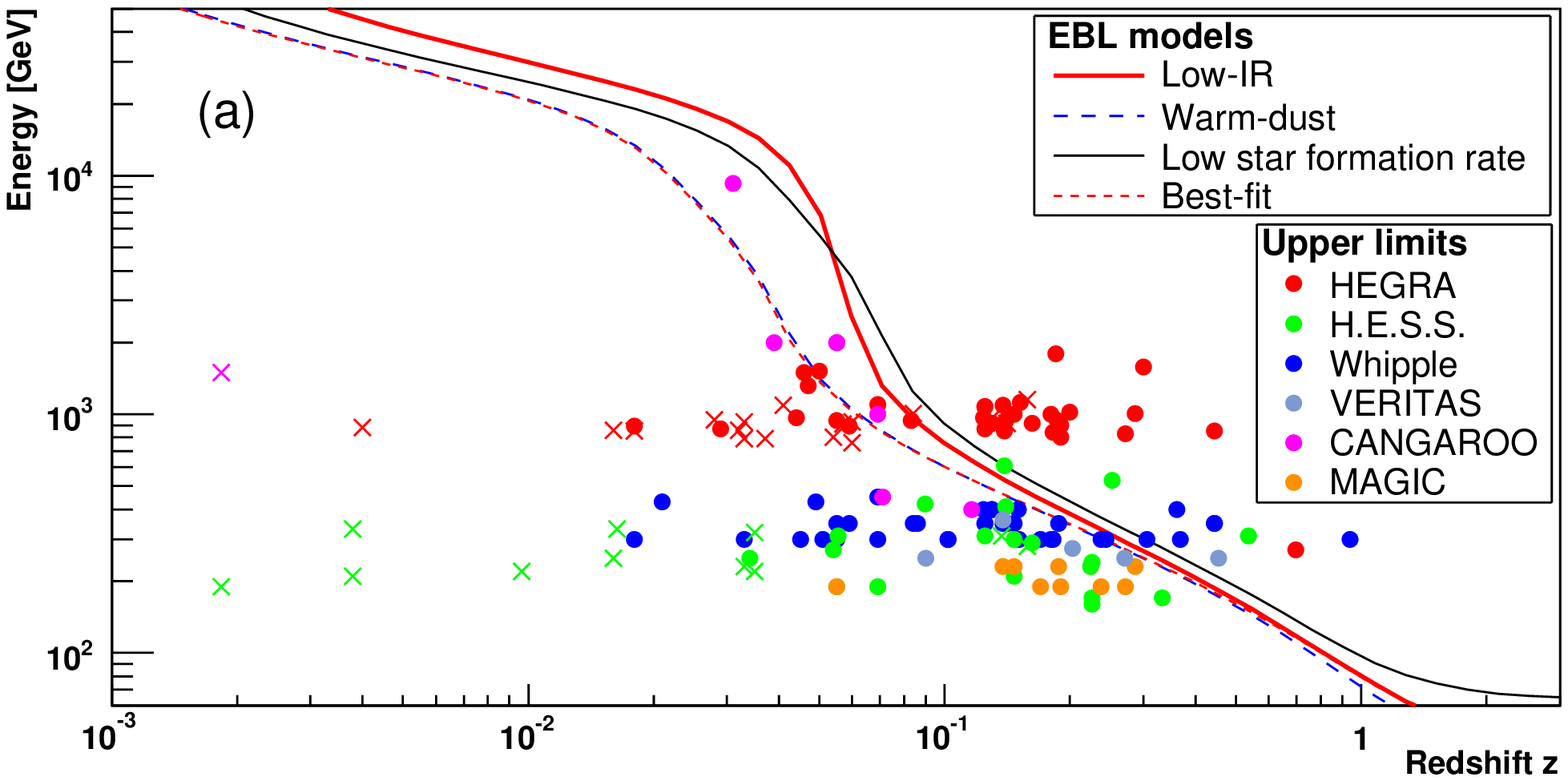}
\includegraphics[width=0.495\linewidth]{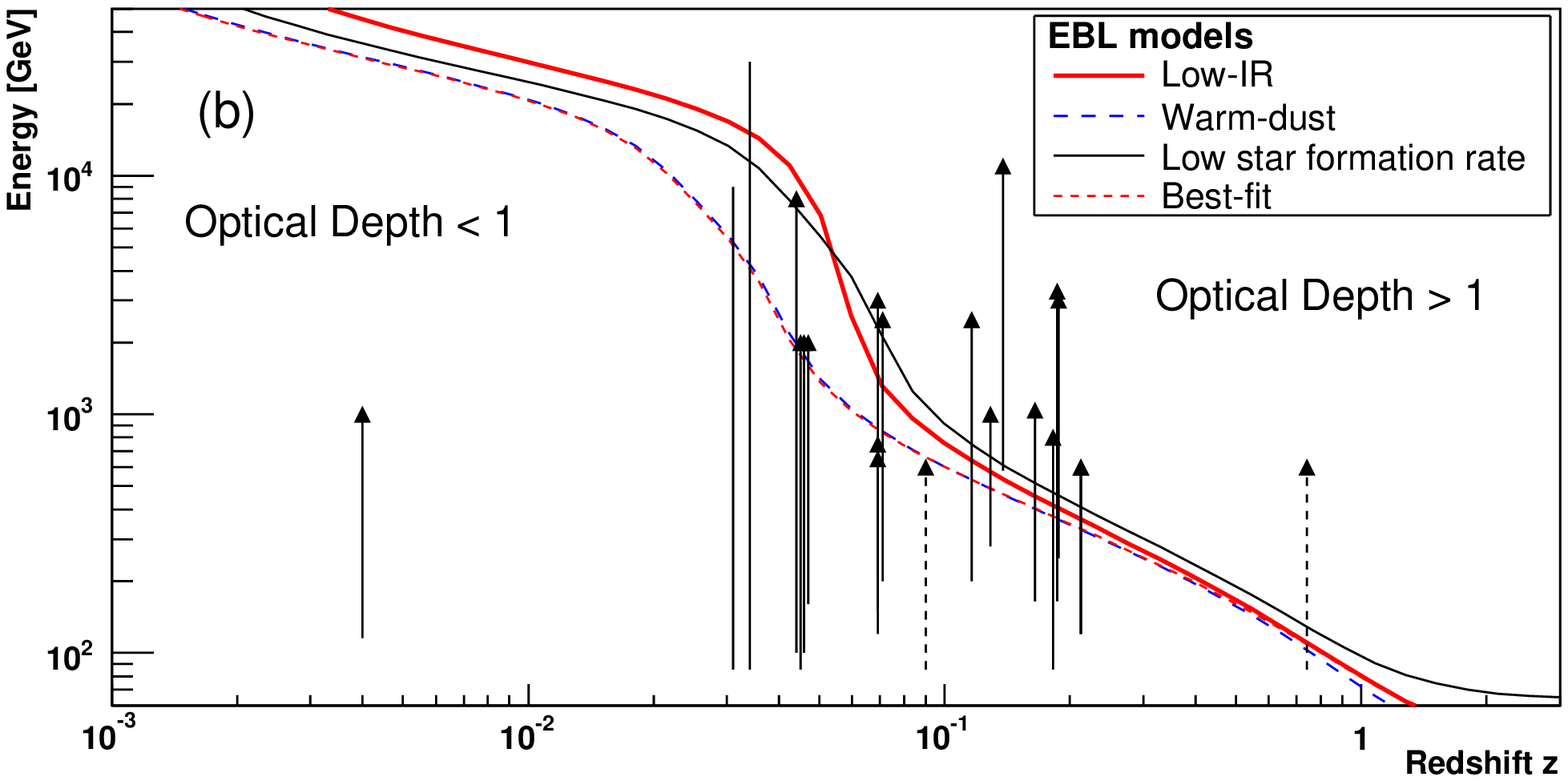}
\caption{{(a)} Lower energy thresholds from searches for VHE
$\gamma$-ray emission from AGN. Dots: blazars, crosses: other AGN
types (starburst galaxies, radio galaxies, etc.). The curves
represent Fazio--Stecker relations (flux attenuation by a factor
$\mathrm{e}^{-1}$) for different EBL models given by
\citet{kneiske4}. Data from searches by HEGRA \citep{Tluczykont},
H.E.S.S. \citep{2005A&A...441..465A,HESS-UL-ICRC}, Whipple
\citep{1995ApJ...452..588K,Horan04}, VERITAS
\citep{VERITAS-UL-ICRC,veul},  CANGAROO
\citep{2002PASA...19...26N} and MAGIC \citep{MAGIC-UL-Paper}.
{(b)} Detected VHE $\gamma$-ray sources (see
Tab.~\ref{tab:parvhe} for references). Shown are the energy ranges
of the measured $\gamma$-ray spectra; high-energy cutoffs were
found only for Mkn\,421 and Mkn\,501, for all other sources arrows
represent spectra possibly continuing to higher energies. The
dashed lines represent PG\,1553+113 at the lower and upper limit
for its redshift, $z>0.09$ \citep{Sbar} and $z<0.74$
\citep{HESS1553,MAGIC1553published}.} \label{fig:comp:compa}}
\end{figure*}

When travelling cosmological distances, VHE $\gamma$-rays interact
with the low-energy photons of the EBL \citep[see,
e.g.][]{nikishov,gould,hauser,2005PhR...409..361K}. The
predominant reaction $ \gamma_{\mbox{\scriptsize{\,VHE}}} +
\gamma_{\mbox{\scriptsize EBL}} \rightarrow
\mathrm{e}^{+}\,\mathrm{e}^{-}$ modifies source-intrinsic
$\gamma$-ray energy spectra. The cross-section of this process
peaks strongly at $E_\mathrm{CM}=1.8\times 2m_\mathrm{e} c^2$,
therefore a given VHE photon energy probes a narrow range of the
EBL spectrum. The part of the EBL to which VHE $\gamma$-rays are
sensitive comprises the (redshifted) relic emission of galaxies
and star-forming systems and the light absorbed and re-emitted by
dust. The EBL attenuation results in a maximum distance over which
photons with a particular energy can survive: The Fazio-Stecker
relation \citep*[FSR,][]{fs70,sjs92} describes the distance at
which the optical depth for a VHE photon of a given energy reaches
unity (attenuation by a factor $\mathrm{e}^{-1}$). 
Thus the FSR defines the {\it cosmological $\gamma$-ray horizon}.
Fig.~\ref{fig:comp:compa}a shows the instrumental low-energy
thresholds from searches for VHE $\gamma$-ray emission from AGN.
Along with these, the FSR for different EBL models as given by
\citet{kneiske4} is plotted. The models differ in the IR density,
dust properties and the star formation rate in the early Universe.
The FSR divides the plot into a region from which no $\gamma$-rays
can reach the Earth and into another region, in which positive
detections are to be expected or a too weak source-intrinsic
emission made detections fail (due to insufficient instrumental
sensitivity). Obviously, with a decreasing instrumental energy
threshold, the visible Universe `opens up', providing access to a
larger source population. Current EBL models result in a
steepening of the intrinsic spectra from $\approx
200\,\mathrm{GeV}$ on (power-law spectra are softened, but their
shape is approximately retained), while for lower energies the
effects are minimal. When an optical depth of one is reached, a
quasi-exponential cutoff in the observed spectra occurs.
Fig.~\ref{fig:comp:compa}b shows the energy ranges over which
blazars in VHE $\gamma$ radiation have been detected. Up to now,
only for the strong, close-by blazars Mkn~421 and Mkn~501
indications of the expected exponential high energy cutoff have
been observed thanks to high $\gamma$ statistics
\citep{2001A&A...366...62A,2001ApJ...560L..45K,MAGIC421published}.
The observed spectra of all other blazars can be accurately
described by power-laws or broken power-laws. While most of the
nearby VHE blazars cannot constrain the current EBL models, some
of the sources at $z>0.1$ start challenging these
\citep{HESSAGNNature,0347,279}, as no cutoffs have been observed
so far at the high-energy ends of their $\gamma$-ray spectra.

\section{Black hole masses in blazars}
\label{sect:comp:bhmavg}
It is well established that all galaxies with a massive bulge
component host supermassive black holes in their centres
\citep{Richstone-98,Bender-03}. There are a couple of indirect
methods to infer the masses of the central BHs: One is to estimate
$M_\bullet$ using the correlation between $M_\bullet$ and the
central velocity dispersion $\sigma$ of the host galaxy
\citep[$M_\bullet-\sigma$ relation,][]{2000ApJ...539L...9F,Gebhardt} found
from stellar and gas kinematics and maser emission. I estimated
the black hole masses of VHE $\gamma$-ray emitting blazars using
the $M_\bullet-\sigma$ relation given by \citet{2002ApJ...574..740T}. This
approach assumes that AGN host galaxies are similar to non-active
galaxies. The velocity dispersions were collected from the
literature or are inferred from the fundamental plane
\citep{1987ApJ...313...59D}, a relation between $\sigma$, the
effective galaxy radius $R_\mathrm{e}$, and the corresponding
surface brightness $\langle\mu_\mathrm{e}\rangle$, which is valid
for elliptical galaxies, in particular also for AGN and radio
galaxies \citep{Bettoni}. Whenever more than one $\sigma$ value is
given in the literature, individual masses were derived for each
of the $\sigma$ values and averaged. The $\sigma$ values and the
resulting black hole masses for the VHE $\gamma$-ray blazars
studied here are given in Tab.~\ref{tab:bhmasses}.

\begin{table*}
\begin{minipage}{147mm}
\caption{Measured velocity dispersions and resulting estimated
black hole masses for the VHE blazars.} \label{tab:bhmasses}
\begin{tabular}{lccccccc}
\hline Object & $\sigma\,[\mathrm{km}\,\mathrm{s}^{-1}]$ &
$\log(M_\bullet/\mathrm{M}_{\sun})$ &
$\sigma\,[\mathrm{km}\,\mathrm{s}^{-1}]$ &
$\log(M_\bullet/\mathrm{M}_{\sun})$ &
$\sigma\,[\mathrm{km}\,\mathrm{s}^{-1}]$ &
$\log(M_\bullet/\mathrm{M}_{\sun})$ &
$\log(M_\bullet/\mathrm{M}_{\sun})$ \\
&Reference 1 &  &  Reference 2 & & Reference 3 & & averaged \\
\hline
Mkn~421         & $219\pm11$ & $8.29\pm0.18$ & $236\pm10$ & $8.42\pm0.15$ & $324\pm18$ & $8.97\pm0.12$ & 8.56 \\
Mkn~501         & $372\pm18$ & $9.21\pm0.11$ & $291\pm13$ & $8.78\pm0.11$ & \nodata    & \nodata       & 9.00 \\
1ES\,2344+514   & $294\pm24$ & $8.80\pm0.15$ & \nodata    & \nodata       & $389\pm20$ & $9.29\pm0.12$ & 9.04 \\
Mkn~180         & $209\pm11$ & $8.20\pm0.19$ & $244\pm10$ & $8.47\pm0.14$ & $251\pm16$ & $8.52\pm0.15$ & 8.40 \\
1ES\,1959+650   & \nodata    & \nodata       & $195\pm15$ & $8.08\pm0.23$ & $219\pm15$ & $8.28\pm0.19$ & 8.18 \\
BL~Lacertae     & \nodata    & \nodata       & \nodata    & \nodata       & $245\pm16$ & $8.48\pm0.15$ & 8.48 \\
PKS~0548-322    & $202\pm24$ & $8.14\pm0.24$ & \nodata    & \nodata       & \nodata    & \nodata       & 8.14 \\
PKS~2005-489    & \nodata    & \nodata       & \nodata    & \nodata       & $257\pm16$ & $8.57\pm0.14$ & \nodata \\
H\,1426+428     & \nodata    & \nodata       & \nodata    & \nodata       & $269\pm16$ & $8.65\pm0.13$ & \nodata \\
H\,2356-309     & \nodata    & \nodata       & \nodata    & \nodata       & $195\pm14$ & $8.08\pm0.23$ & \nodata \\
1ES\,0229+200   & \nodata    & \nodata       & \nodata    & \nodata       & $363\pm19$ & $9.16\pm0.11$ & \nodata \\
1ES\,1218+304   & \nodata    & \nodata       & \nodata    & \nodata       & $191\pm14$ & $8.04\pm0.24$ & \nodata \\
1ES\,0347-121   & \nodata    & \nodata       & \nodata    & \nodata       & $214\pm15$ & $8.24\pm0.19$ & \nodata \\
1ES\,1011+496   & \nodata    & \nodata       & \nodata    & \nodata       & $219\pm15$ & $8.28\pm0.19$ & \nodata \\
\hline
\end{tabular}

\medskip
References: (1) \citet*{Barth2003}; (2)
\citet*{2002ApJ...569L..35F}; (3) \citet*{2002AA...389..742W}. The
velocity dispersions $\sigma$ were translated into estimated BH
masses using the $M_\bullet-\sigma$ relation from
\citet{2002ApJ...574..740T}. The $\sigma$ values taken from
\citet{2002AA...389..742W} were indirectly determined using the
fundamental plane of radio galaxies \citep{Bettoni}. BH masses
given in units of the solar mass, $\mathrm{M}_{\sun}$. If more
than one $M_\bullet$ value is given, the average $M_\bullet$ is
used. Due to the possible different systematic errors of the
individual data sets, the largest error was assumed as error of
the average $M_\bullet$.
\end{minipage}
\end{table*}
The determination of $M_\bullet$ suffers from rather large
systematic uncertainties due to the different methods used to
derive $\sigma$. The relation between $M_\bullet$ and bulge
luminosity $L_\mathrm{B}$ \citep{1995ARAA..33..581K} has generally
a larger scatter than the $M_\bullet-\sigma$ relation and was therefore
only used for PKS\,2155-304, because for this blazar no $\sigma$
or $\langle\mu_\mathrm{e}\rangle$ measurement is available. I used
the $R$-band luminosity given by \citet{1996MNRAS.283..241F} to
calculate $\log(M_\bullet/\mathrm{M}_{\sun})=8.98\pm0.44$ using
eq.~12 in \citet{2007MNRAS.379..711G}. \citet{2155flare} give an
estimate of $M_\bullet=(1\dots2)\times10^9 \mathrm{M}_{\sun}$. For
a comparative study using values inferred by different methods is
not advisable due to possible different systematics. Therefore,
the $M_\bullet$ value used for PKS\,2155-304 should be taken with
care in the following. The BH mass of 3C~279 was determined using
the virial BH mass estimate of \citet{McLure} and is given as
$\log(M_\bullet/\mathrm{M}_{\sun}) = 8.912$ by
\citet{2001MNRAS.327.1111G}. 
For two of the blazars under study, PG\,1553+113 and
1ES\,1101-232, no $M_\bullet$ estimations exist yet.

\begin{figure*}
\center{\includegraphics[width=0.65\linewidth]{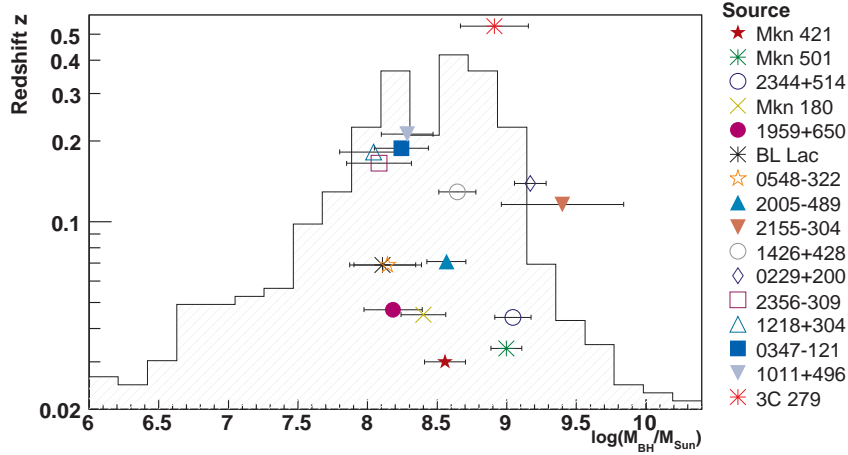}
\caption{Redshift vs. $M_\bullet$ distribution for the known VHE
$\gamma$-ray emitting AGN. The superimposed histogram shows the
BH mass distribution of the 375 AGN given in
Fig.~\ref{fig:comp:compazwu} (linear vertical scale independent of
$z$). \label{fig:comp:compaz}}}
\end{figure*}

\begin{figure*}
\center{\includegraphics[width=0.65\linewidth]{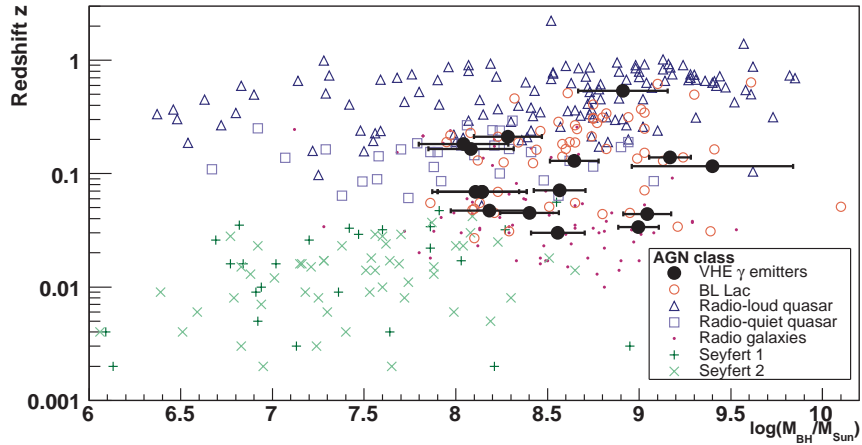}
\caption{The redshift vs. $M_\bullet$ distribution for 375 AGN
collected by \citet{2002ApJ...579..530W} and the known VHE
$\gamma$-ray emitting AGN.\label{fig:comp:compazwu}}}
\end{figure*}

Recent estimations of the BH masses for 452 AGN find them
distributed over a large range of
$(10^6-7\times10^{9})\,\mathrm{M}_{\sun}$ with no evidence for
dependencies on the radio loudness of the objects
\citep{2002ApJ...579..530W,2002ApJ...581L...5W}. A recent study of
the BH mass distribution of 66 BL~Lac objects
\citep{2005ApJ...631..762W} reports an $M_\bullet$ range of
$(10^7-4\times10^{9})\,\mathrm{M}_{\sun}$ and could also not find
a correlation of $M_\bullet$ with radio or X-ray luminosity. (In
Sect. \ref{sect:intrinsicspectra} the distribution of VHE blazars
in luminosity and $M_\bullet$ is discussed). As
Fig.~\ref{fig:comp:compaz} shows, there is no dependence of the BH
masses of the VHE blazars on their redshift, but they are rather
flatly distributed in their BH masses between $(10^8-10^{9.5})
\mathrm{M}_{\sun}$.
Although AGN harbour BHs with $M_\bullet > 10^6
\mathrm{M}_{\sun}$, up to now only blazars with rather massive
BHs, $M_\bullet \ga 10^8 \mathrm{M}_{\sun}$, have been discovered
in VHE $\gamma$-rays, raising the question whether a physics
reason is responsible for the non-detection of blazars with less 
massive BHs in the mass range $(10^7-10^8)\,\mathrm{M}_{\sun}$. 
There exist studies that find radio-loud AGN, and therefore also
blazars, to be associated with BHs with $M_\bullet \ga 10^9
\mathrm{M}_{\sun}$ \citep{laor2000} or at least on average with
more massive BHs than radio-quiet AGN \citep{mm06}. The latter
authors also report a threshold BH mass for the onset of radio
activity, with very little dependence of the radio output on the
BH mass once above the threshold mass. Whether such a mass
threshold is also at work for the VHE emission, remains subject
for further studies at this point.
The BH masses of the VHE blazars are compared to those of 375 AGN
collected by \citet{2002ApJ...579..530W} in
Fig.~\ref{fig:comp:compazwu}. The confinement of Seyfert galaxy
measurements to low redshifts presumably is due to a selection
effect: These are spiral galaxies and therefore expected to
harbour comparatively low-mass BHs. Distant Seyfert galaxies
($z\ga1.0$) might just not be luminous enough to obtain
$M_\bullet$ measurements. Conversely, quasars are too rare as to
be found in small volumes and thus at small distances.

\section{Intrinsic VHE $\gamma$-ray emission parameters}
\label{sect:intrvhe}
VHE $\gamma$-ray observations enable us to look deep into the
emission regions of blazar jets and thus convey information on the
responsible particle acceleration and cooling processes. Here I
study primarily the differential energy spectra in the VHE domain,
which are summarised in  Tab.~\ref{tab:parvhe}.
\begin{table*}
\begin{minipage}{175mm}
\caption{Measured VHE blazar spectra, reconstructed intrinsic
spectral indices and luminosities.} \label{tab:parvhe}
\begin{tabular}{lllcccc}
\hline Object & Measured Energy Spectrum $\mathrm{d}F/\mathrm{d}E$
& Reference &
Intrinsic & $\nu_\gamma L_\gamma$ \\
 & $[\mathrm{TeV}^{-1}\,\mathrm{cm}^{-2}\,\mathrm{s}^{-1}]$
&  & Slope $\Gamma$ &
$[\mathrm{erg}\,\mathrm{s}^{-1}\,\mathrm{sr}^{-1}]$\\
\hline
Mkn~421       & $(12.1\pm0.5)10^{-12} (E/1.0\,\mathrm{TeV})^{-3.09\pm0.07} $ & \citet{AharonianMkn4211999} & $2.85\pm0.58$ & $(9.68\pm0.40)\times10^{43}$ \\
Mkn~501       & $(8.4\pm0.5)10^{-12} (E/1.0\,\mathrm{TeV})^{-2.76\pm0.08} $ & \citet{Aharonian2001} & $2.49\pm0.84$ & $(6.89\pm0.41)\times10^{43}$ \\
1ES\,2344+514 & $(1.2\pm0.2)10^{-11} (E/0.5\,\mathrm{TeV})^{-2.95\pm0.12} $ & \citet{MAGIC2344published} & $2.67\pm0.21$ & $(2.80\pm0.47)\times10^{43}$ \\
Mkn~180       & $(4.5\pm1.8)10^{-11} (E/0.3\,\mathrm{TeV})^{-3.3\pm0.7} $ & \citet{MAGIC180} & $3.06\pm0.50$ & $(2.04\pm0.81)\times10^{43}$ \\
1ES\,1959+650 & $(3.4\pm0.5)10^{-12} (E/1.0\,\mathrm{TeV})^{-2.72\pm0.14} $ & \citet{NadiaPaper} & $2.37\pm0.29$ & $(5.95\pm0.88)\times10^{43}$ \\
BL Lacertae   & $(1.9\pm0.5)10^{-11} (E/0.3\,\mathrm{TeV})^{-3.64\pm0.54} $ & \citet{MAGICBLLac} & $3.17\pm0.25$ & $(2.35\pm0.62)\times10^{43}$ \\
PKS\,0548-322 & $(1.9\pm0.4)10^{-13} (E/1.0\,\mathrm{TeV})^{-2.8\pm0.3} $ & \citet{0548icrc} & $2.38\pm0.28$ & $(9.66\pm2.05)\times10^{42}$ \\
PKS\,2005-489 & $(1.9\pm0.7)10^{-13} (E/1.0\,\mathrm{TeV})^{-4.0\pm0.4} $ & \citet{HESS2005} & $3.52\pm0.27$ & $(2.53\pm0.93)\times10^{43}$ \\
PKS\,2155-304 & $(1.96\pm0.12)10^{-12} (E/1.0\,\mathrm{TeV})^{-3.32\pm0.06}$ for $E< 700\,\mathrm{GeV}$, & \citet{Aharonian2005} & $2.43\pm0.64$ & $(6.32\pm0.39)\times10^{44}$ \\
              & $(2.4^{+0.4}_{-0.3})10^{-12} (0.7\pm0.2)^{(3.79^{+0.46}_{-0.27}-3.15^{+0.10}_{-0.12})}$ & & &  \\
              & $\times (E/1.0\,\mathrm{TeV})^{-3.79^{+0.46}_{-0.27}}$ for $E> 700\,\mathrm{GeV}$  & &  & \\
H\,1426+428  & $(2.9\pm1.1)10^{-11} (E/0.43\,\mathrm{TeV})^{-2.6\pm0.6} $ & \citet{2001ICRC....7.2622H}, & $1.58\pm0.23$ & $(7.42\pm2.81)\times10^{44}$ \\
& & \citet{2002AA...384L..23A} &   &   \\
1ES\,0229+200 & $(2.34\pm0.37)10^{-14} (E/3.0\,\mathrm{TeV})^{-2.5\pm0.2} $ & \citet{0229} & $1.39\pm0.30$ & $(6.13\pm0.98)\times10^{43}$ \\
H\,2356-309  & $(3.08\pm0.75)10^{-13} (E/1.0\,\mathrm{TeV})^{-3.06\pm0.4} $ & \citet{HESSAGNNature} & $1.77\pm0.37$ & $(3.29\pm0.80)\times10^{44}$ \\
1ES\,1218+304 & $(8.1\pm2.1)10^{-11} (E/0.25\,\mathrm{TeV})^{-3.0\pm0.4} $ & \citet{MAGIC1218} & $1.97\pm0.40$ & $(1.26\pm0.33)\times10^{45}$ \\
1ES\,1101-232 & $(4.44\pm0.74)10^{-13} (E/1.0\,\mathrm{TeV})^{-2.88\pm0.14} $ & \citet{HESSAGNNature} & $1.33\pm0.37$ & $(3.19\pm0.53)\times10^{44}$ \\
1ES\,0347-121 & $(4.52\pm0.85)10^{-13} (E/1.0\,\mathrm{TeV})^{-3.10\pm0.23} $ & \citet{0347} & $1.76\pm0.14$ & $(3.19\pm0.60)\times10^{44}$ \\
1ES\,1011+496 & $(2.0\pm0.1)10^{-10} (E/0.2\,\mathrm{TeV})^{-4.0\pm0.5} $ & \citet{1011} & $2.56\pm0.29$ & $(15.4\pm0.77)\times10^{44}$ \\
PG\,1553+113$^a$  & $(1.8\pm0.3)10^{-10} (E/0.2\,\mathrm{TeV})^{-4.21\pm0.25} $ & \citet{MAGIC1553published} & $3.68\pm0.68$ & $(6.40\pm1.07)\times10^{43}$ \\
PG\,1553+113$^b$  & $(1.8\pm0.3)10^{-10} (E/0.2\,\mathrm{TeV})^{-4.21\pm0.25} $ & \citet{MAGIC1553published} & $2.34\pm0.46$ & $(3.14\pm0.52)\times10^{45}$ \\
\hline
Mkn~421$^c$       & $(23.40\pm0.73)10^{-11} (E/1.0\,\mathrm{TeV})^{-2.32\pm0.03} $ & \citet{Krennrich2002} & $2.09\pm0.30$ & $(1.08\pm0.03)\times10^{45}$ \\
Mkn~501$^c$       & $(2.50\pm0.16)10^{-10} (E/1.0\,\mathrm{TeV})^{-2.22\pm0.04} $ & \citet{AharonianMkn501-1999a} & $1.95\pm0.41$ & $(1.39\pm0.09)\times10^{45}$ \\
1ES\,2344+514$^c$ & $(5.1\pm1.0)10^{-11} (E/1.0\,\mathrm{TeV})^{-2.54\pm0.17} $ & \citet{Schroedter} & $2.20\pm0.31$ & $(6.65\pm1.30)\times10^{44}$ \\
1ES\,1959+650$^c$ & $(1.23\pm0.25)10^{-10} (E/1.0\,\mathrm{TeV})^{-2.78\pm0.12} $ & \citet{2005ApJ...621..181D} & $2.43\pm0.29$ & $(2.25\pm0.46)\times10^{45}$ \\
PKS\,2155-304$^c$ & $(2.06\pm0.16)10^{-10}
(E/1.0\,\mathrm{TeV})^{-2.71\pm0.06}$ for $E< 340\,\mathrm{GeV}$, & \citet{2155flare} & $2.28\pm0.40$ & $(2.74\pm0.17)\times10^{46}$ \\
              & $(2.06\pm0.16)10^{-10} (0.430\pm0.022)^{(3.53\pm0.05)-(2.71\pm0.06)}$  & &  &  \\
              & $\times (E/1.0\,\mathrm{TeV})^{-3.53\pm0.05}$ for $E> 340\,\mathrm{GeV}$  & &  & \\
\hline
\end{tabular}

\medskip
$\Gamma$ denotes the reconstructed (intrinsic) VHE spectral
power-law index at 500 GeV and $\nu_\gamma L_\gamma$ represents
the source luminosity at 500 GeV. Both values were calculated from
the measured spectra assuming a \citeauthor{kneiske4} `low-IR' EBL
density. $^a$ at an assumed $z=0.1$.~$^b$ at an assumed
$z=0.3$.~$^c$ spectrum measured during a flare state of the
respective blazar.
\end{minipage}
\end{table*}
For the sources Mkn~421, Mkn~501, PKS\,2155-304, 1ES\,1959+650,
and 1ES\,2344+514 observations of clearly distinct flux states
exist. Accordingly, for each of those sources two spectra, one
`low-state' and one `high-state' spectrum, are considered.
Low-state spectra are characterised by the absence of high ($\ga
0.2$~Crab units\footnote{The Crab nebula exhibits a strong,
constant VHE $\gamma$-ray flux and is therefore often considered a
standard candle in VHE $\gamma$-ray astronomy}) flux levels and
short-term variability (probably beyond instrumental sensitivity
though), while flare spectra were obtained during outbursts of the
respective sources ({\em viz.} the 1995 December 20 flare of
1ES\,2344+514, the 1997 flare of Mkn~501, the 2002 flare of
1ES\,1959+650 and the 2006 July 28 flare of PKS\,2155-304). At
present for none of these sources, even for those with only one
flux state, can a true baseline flux state be claimed, although
low-flux states have been observed
\citep{NadiaPaper,MAGIC2344published}. Long-term monitoring
campaigns are currently performed to address this issue
\citep{magicmonitoring,veritasmonitoring,hessmonitoring}.

The measured spectra suffer $\gamma\gamma$ absorption on photons
of the EBL as shown in Sect.~\ref{sect:grhorsect}. The intrinsic
source spectra are reconstructed employing \citep{DanielDiplom}
the EBL `low-IR' model given in \citet{kneiske4}, which assumes
the least possible infrared star formation rate as allowed by
galaxy counts and which is in reasonable agreement with other
models \citep*{primack421,steckerNEW}. Note that due to the fact
that the VHE $\gamma$-rays are attenuated exponentially with the
optical depth, an accurate knowledge of the EBL is crucial for the
individual interpretation of the intrinsic VHE $\gamma$-ray
spectra.

In the following, I will use two observables to characterise the
VHE $\gamma$-ray emission: The $K$-corrected \citep{hogg-02}
luminosity at $500\,\mathrm{GeV}$, $\nu_\gamma L_{\gamma} = 4 \pi
d_\mathrm{L}^2 \cdot (500\,\mathrm{GeV})^2
F(500\,\mathrm{GeV}/(1+z))/(1+z)$ with the luminosity distance
$d_\mathrm{L}$ and the intrinsic photon index $\Gamma$ in the
region around 500~GeV, which is determined by fitting the
intrinsic spectra with pure power-laws of the form
$\mathrm{d}F/\mathrm{d}E = f_0 \cdot (E/E_0)^{-\Gamma}$. These two
parameters act as proxies for the peak position and the spectral
shape on the falling edge of the high-energy bump, which cannot,
as explained before, easily be determined from the existing VHE
data. For extraction of the luminosity and of the spectral slope,
the region around 500 GeV was chosen because all blazars under
study have measured spectra in this energy region. All
calculations and fits have been performed in energy ranges where
$\gamma$-ray spectra for the respective blazars have actually been
observed, so that no extrapolations in energy regions not covered
by the data were required. All in all, extragalactic source
observations included in this paper cover the energy range
$85\,\mathrm{GeV} \leq E \leq 11 \,\mathrm{TeV}$. For the
determination of the luminosity distances the cosmological
parameters given in \citet{WMAPpaperpublished} were used:
$\Omega_\mathrm{m}h^2 = 0.127^{+0.007}_{-0.013}$;
$\Omega_\mathrm{b}h^2 = 0.0223^{+0.0007}_{-0.0009}$ with the
Hubble constant $H_0=100\cdot
h\,\mathrm{km}\,\mathrm{s}^{-1}\,\mathrm{Mpc}^{-1} =
73^{+3}_{-3}\,\mathrm{km}\,\mathrm{s}^{-1}\,\mathrm{Mpc}^{-1}$.

Tab.~\ref{tab:parvhe} also shows the resulting source luminosities
in the VHE region and the spectral slope of the reconstructed
source-intrinsic spectra. While the luminosities range from
$\approx 10^{43}\,\mathrm{erg}\,\mathrm{s}^{-1}\,\mathrm{sr}^{-1}$
to $\approx
3\times10^{45}\,\mathrm{erg}\,\mathrm{s}^{-1}\,\mathrm{sr}^{-1}$
($\approx
10^{45}\,\mathrm{erg}\,\mathrm{s}^{-1}\,\mathrm{sr}^{-1}$ to
$\approx
3\times10^{46}\,\mathrm{erg}\,\mathrm{s}^{-1}\,\mathrm{sr}^{-1}$
for blazars in outburst), the photon indices of the reconstructed
intrinsic spectra vary between $\Gamma=1.4 - 3.3$, except for
1ES\,1101-232, which probably has an intrinsic spectrum peaking
far beyond $E=1$\,TeV \citep{1101}. The VHE measurements lie close
to, but generally above the maximum of the high-energy bump (which
occurs at $\Gamma=2$). Spectra with $\Gamma<1.5$ are difficult to
explain in current acceleration models
(\citealt{2001RPPh...64..429M,HESSAGNNature}; but see
\citealt{SS3,SS4,sbs07} for models that would explain harder
spectra). The rather hard intrinsic slopes found for some sources
go in line with indications that the EBL absorption effects are
still smaller than currently modelled
\citep[e.g.][]{HESSAGNNature}: a lower EBL level would soften the
intrinsic spectra inferred here, i.e. increase the value of
$\Gamma$.

Before instruments like MAGIC and H.E.S.S. became operational, the
average observed photon index in BL~Lac objects was
$\Gamma\approx2.3$. This raised the expectation that AGN would in
general exhibit rather hard spectra in the VHE range, which also
would be compatible with the average EGRET blazar spectrum (at
MeV-GeV energies), found to have a slope of
$\Gamma_\mathrm{EGRET}=2.27$ \citep{2007ApJ...666..128V}. In fact,
the average spectral slope at VHE has not changed much, although
the scatter of $\Gamma$ increased as a new population of objects
with intrinsically rather soft spectra (Mkn\,180, PKS\,2005-489)
has been tapped, and at the same time distant, hard-spectrum
sources were found.

For PG\,1553+113 with its unknown redshift (see
Sect.~\ref{sect:comp:501lumilim} for details), two possible
distances ($z=0.1$, $z=0.3$) were assumed in this paper. The
resulting `intrinsic spectra', however, are only given for
illustrative purposes and are not used for any conclusions
throughout this study unless stated otherwise.

\section{Correlation studies}
\label{sect:corstudies}
\subsection{Correlation of X-ray, optical, radio and VHE $\gamma$-ray luminosity}
\label{sect:corstudies1}
\begin{figure*}
\center{\includegraphics[width=.65\linewidth]{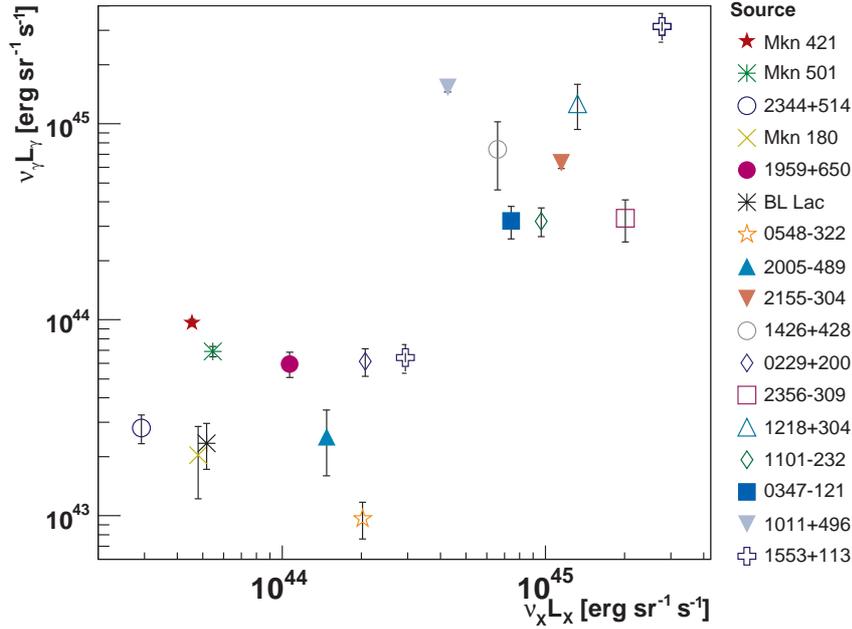}
\caption{VHE $\gamma$-ray luminosity $\nu_\gamma L_\gamma$ vs.
X-ray luminosity at 1 keV, $\nu_\mathrm{X} L_\mathrm{X}$ for 17
VHE blazars. The two data points for PG\,1553+113 (open crosses)
are for assumed redshifts of $z=0.1$ ($\nu_\gamma
L_{\gamma}=6.4\times10^{43}\,\mathrm{erg}\,\mathrm{sr}^{-1}\,\mathrm{s}^{-1}$)
and $z=0.3$ ($\nu_\gamma
L_{\gamma}=3.1\times10^{45}\,\mathrm{erg}\,\mathrm{sr}^{-1}\,\mathrm{s}^{-1}$),
respectively, and are not used in the fit and for determining the
correlation coefficient (see text). It should be noted that VHE
data points have an additional systematic error of typically 35
per cent. The systematic error of the X-ray luminosities is
unfortunately unknown.\label{fig:comp:compaxrc}}}
\end{figure*}

In SSC models, the X-ray and the VHE emission are closely
connected, owing to their common origin. While in some blazars
clear evidence for a corresponding correlation has been observed
\citep[Mkn 421,][]{krawczinski,bla421,MAGIC421published}, the
connection is only weak for other ones \citep[Mkn
501,][]{MAGIC-501} or even non-existing during some flare states
\citep[the 1ES 1959+650 {\it orphan flare}
case,][]{2005ApJ...621..181D}. Fig.~\ref{fig:comp:compaxrc} shows
$\nu_\gamma L_\gamma$  versus the X-ray luminosity at 1 keV
($\nu_\mathrm{X} L_\mathrm{X}$; from \citealt{costamante}). Note
that high thermal contributions at 1 keV are unlikely and would
imply a very high amount of gas and pressure. Unfortunately, the
X-ray and VHE data have not been taken simultaneously.
While VHE measurements during outbursts were not used in the
Figure, variations in the X-ray domain for the blazars under study
are, according to the compilation of X-ray fluxes in
\citet{2001AA...375..739D}, not larger than a factor 4.6 (most
extreme object: Mkn\,501) with an average of a factor 1.4 and a
variance of 2.2. According to expectations, a trend towards a
correlation is visible. When including all data points except for
those representing PG\,1553+113, I find a correlation coefficient
of $r=0.76^{+0.09}_{-0.14}$, which is within 3.6 standard
deviations different from zero. A linear fit to the data yields a
slope of $m=1.11\pm0.09$ ($\chi^2_\mathrm{red} = 61.8/14$).

Recently, optical triggers lead to the successful discoveries of
the VHE blazars Mkn~180 \citep{MAGIC180} and 1ES\,1011+496
\citep{1011}. At times of lower optical emission the latter in
fact  showed a lower flux and thus only yielded a marginal
detection \citep{MAGIC-UL-Paper}. This seems to imply the
existence of a VHE -- optical connection, which would be a
convenient proxy for finding new VHE blazars. However, in the past
no \citep{bla421,2006ApJ...641..740R,MAGIC1553published} or only
weak \citep{1996ApJ...472L...9B} evidence for optical or radio to
VHE correlations was found for individual observations of Mkn~421
and PG\,1553+113.

\begin{figure*}
\center{\includegraphics[width=\linewidth]{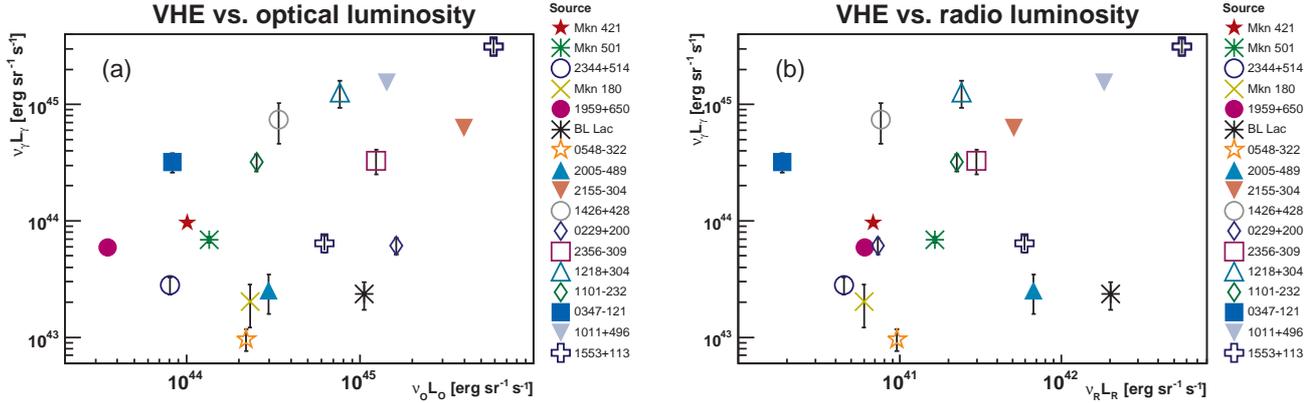}
\caption{{(a)} Correlation of VHE luminosity $\nu_\gamma
L_\gamma$ and optical luminosity $\nu_\mathrm{O} L_\mathrm{O}$.
{(b)} Correlation of VHE luminosity $\nu_\gamma L_\gamma$ and
radio luminosity $\nu_\mathrm{R} L_\mathrm{R}$. The two data
points for PG\,1553+113 (open crosses) are for assumed redshifts
of $z=0.1$ ($\nu_\gamma
L_{\gamma}=6.4\times10^{43}\,\mathrm{erg}\,\mathrm{sr}^{-1}\,\mathrm{s}^{-1}$)
and $z=0.3$ ($\nu_\gamma
L_{\gamma}=3.1\times10^{45}\,\mathrm{erg}\,\mathrm{sr}^{-1}\,\mathrm{s}^{-1}$),
respectively. \label{fig:comp:VHE_vs_O}}}
\end{figure*}

The optical flux data at 5500~\AA\, used here were collected from
several source catalogues by \citet{costamante}, as were the radio
flux data at 5~GHz. I converted them into luminosities taking into
account the appropriate luminosity distances $d_\mathrm{L}$.
Fig.~\ref{fig:comp:VHE_vs_O} shows the corresponding correlations:
the data in the VHE -- optical plane feature a larger scatter than
that in the VHE -- X-ray data, while in the VHE -- radio plane no
clear trend is seen.

\begin{figure*}
\center{\includegraphics[width=\linewidth]{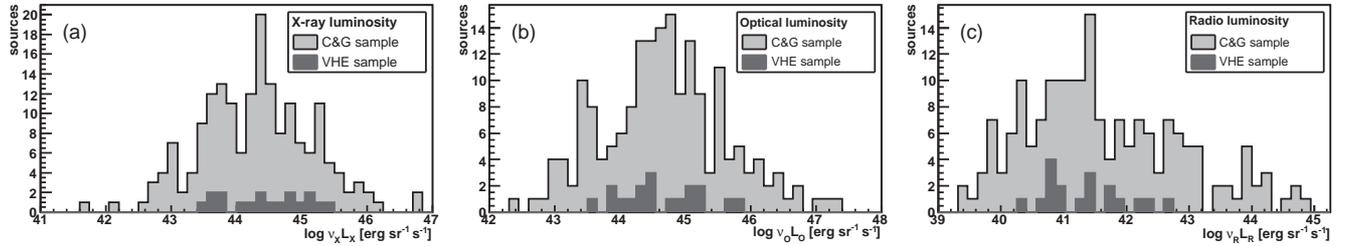}
\caption{Histograms of {(a)}  1 keV X-ray, {(b)} 5500
\AA\, optical and {(c)} 5 GHz radio luminosity of the VHE blazars and all
those blazars considered by \citet{costamante} with determined redshifts (183
out of the total set of 246 blazars).
\label{fig:comp:CG_lumiHistograms}}}
\end{figure*}

VHE blazars might populate only restricted ranges in X-ray,
optical, or radio luminosity distributions. To test this, I use
X-ray, optical and radio data from the full set of 246 sources
considered by \citet{costamante}. I rejected the sources for which
no redshift is known and converted the remaining 183 fluxes into
luminosities. These blazars are compared to the VHE blazars in
Fig. \ref{fig:comp:CG_lumiHistograms}. In none of the three
distributions can substantial deviations of the VHE blazars from
the overall set of blazars be found.

\subsection{Correlations of intrinsic photon index and VHE luminosity with the synchrotron peak frequency}

\begin{figure*}
\center{\includegraphics[width=\linewidth]{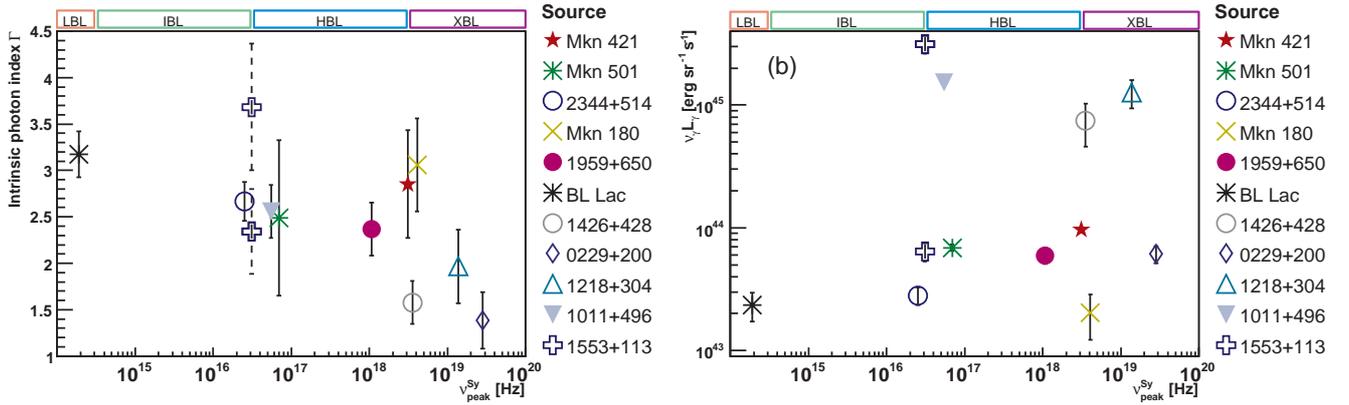}
\caption{{(a)} Photon index $\Gamma$ vs. synchrotron peak
frequency $\nu^\mathrm{Sy}_\mathrm{peak}$ and {(b)} VHE
luminosity vs. synchrotron peak frequency
$\nu^\mathrm{Sy}_\mathrm{peak}$ for eleven VHE blazars. The
synchrotron peak frequencies and the LBL-IBL-HBL classification
are taken from \citet{Nieppola}. The two data points for
PG\,1553+113 (open crosses) are for assumed redshifts of $z=0.1$
($\Gamma=3.68\pm0.68$; $\nu_\gamma
L_{\gamma}=6.4\times10^{43}\,\mathrm{erg}\,\mathrm{sr}^{-1}\,\mathrm{s}^{-1}$)
and $z=0.3$ ($\Gamma=2.34\pm0.46$; $\nu_\gamma
L_{\gamma}=3.1\times10^{45}\,\mathrm{erg}\,\mathrm{sr}^{-1}\,\mathrm{s}^{-1}$),
respectively, and are not used in the fit and for determining the
correlation coefficient in the left Figure {(a)}.\label{fig:comp:VHE_vs_Npeak}}}
\end{figure*}

\begin{figure*}
\center{\includegraphics[width=0.66\linewidth]{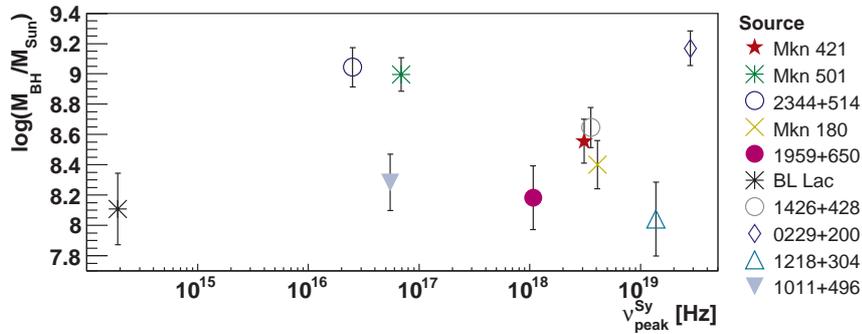}
\caption{Estimated BH mass vs. synchrotron peak frequency
$\nu^\mathrm{Sy}_\mathrm{peak}$ for those VHE blazars with both
quantities known. \label{fig:comp:MBH_vs_Npeak}}}
\end{figure*}

In SSC models the VHE peak, identified with the inverse Compton
(IC) peak with a maximum at $\nu^\mathrm{IC}_\mathrm{peak}$,
resembles the form \citep[e.g. ][]{fossati} of the synchrotron
peak at $\nu^\mathrm{Sy}_\mathrm{peak}$, displaced by the squared
Lorentz factor $\nu^\mathrm{IC}_\mathrm{peak} /
\nu^\mathrm{Sy}_\mathrm{peak} \sim \gamma^2$
\citep{Tavecchio1998}. \citet*{Nieppola} collected
(non-simultaneous) multiwavelength data for a large ($>300$) set
of blazars, which includes eleven of the VHE blazars under study
here. The data, covering frequencies from the 5 GHz radio to the
2.4 keV X-ray domain, were used by \citeauthor{Nieppola} to
reconstruct the SEDs of these blazars and to infer the synchrotron
peak frequency $\nu_\mathrm{peak}$. I test for correlations
between $\nu^\mathrm{Sy}_\mathrm{peak}$ and the VHE luminosity and
spectral slope (Fig.~\ref{fig:comp:VHE_vs_Npeak}). According to
expectations from SSC models, I find a correlation of the photon
index $\Gamma$ with $\nu^\mathrm{Sy}_\mathrm{peak}$. The
correlation coefficient is $r=-0.63^{+0.28}_{-0.18}$ and a linear
fit to the data as shown in Fig.~\ref{fig:comp:VHE_vs_Npeak}a
yields a $\chi^2_\mathrm{red}=12.1/8$.
Fig.~\ref{fig:comp:VHE_vs_Npeak}b shows the corresponding data
points in the $\nu_\gamma L_\gamma -
\nu^\mathrm{Sy}_\mathrm{peak}$ plane, in which no correlation is
apparent. Finally, in Fig.~\ref{fig:comp:MBH_vs_Npeak}, I check
whether the LBL--HBL transition of the VHE blazars is connected
with the BH masses of their host galaxies. Ten of the VHE blazars,
for which both $\nu^\mathrm{Sy}_\mathrm{peak}$ and $M_\bullet$
measurements are available, show no trend towards a correlation.

\subsection{Correlation between the intrinsic photon index and the VHE $\gamma$-ray
luminosity}
\begin{figure}
\center{\includegraphics[width=\linewidth]{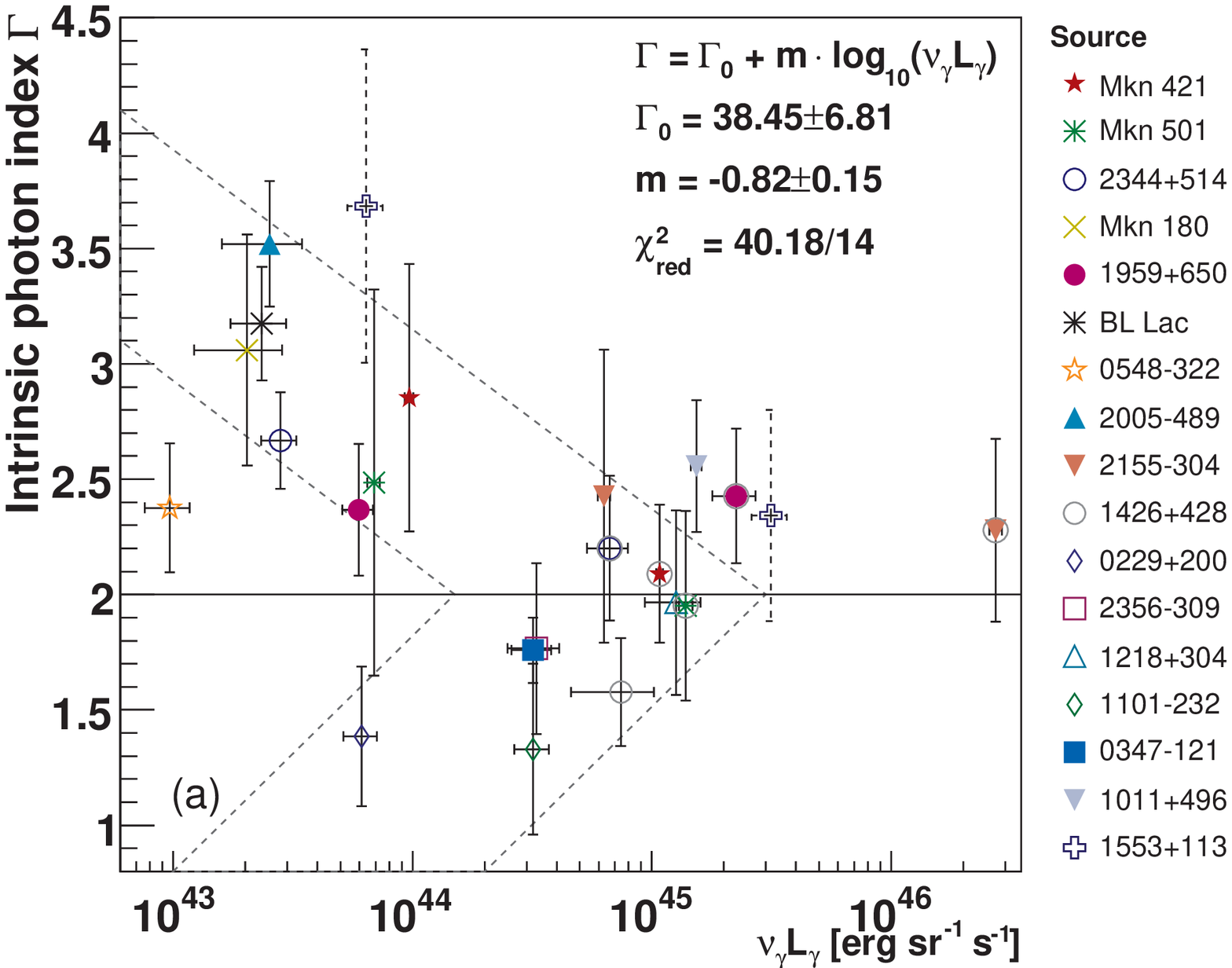}
\includegraphics[width=\linewidth]{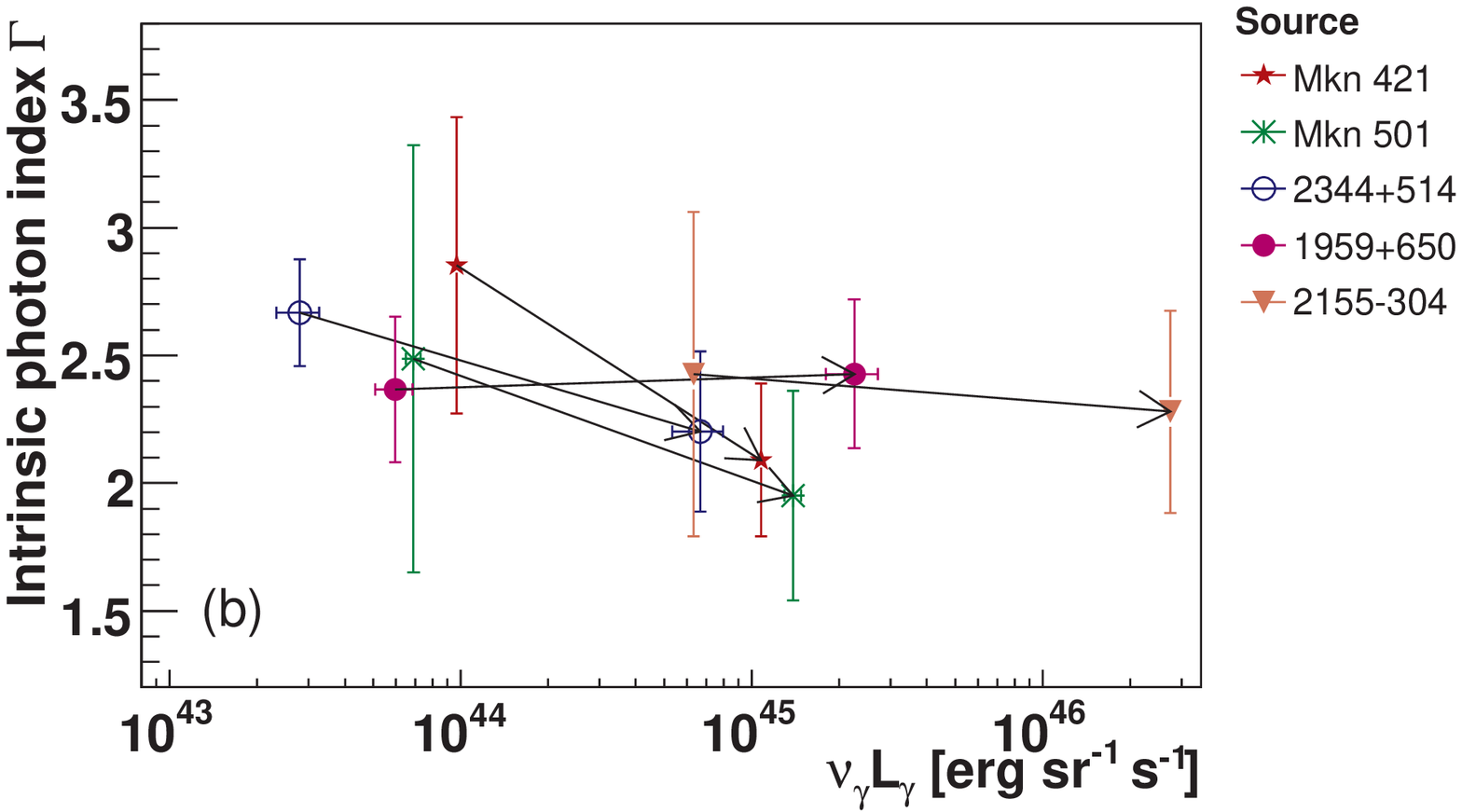} \caption{{(a)} Intrinsic photon
index vs. luminosity. Additional flare states of sources are
marked by grey circles. The results of a linear fit of the form
$\Gamma=\Gamma_0+m \log_{10} (\nu_\gamma L_\gamma)$ are given in
the figure. The two data points for PG\,1553+113 (open crosses),
not included in this fit, are for assumed redshifts of $z=0.1$
(low luminosity) and $z=0.3$ (high luminosity), respectively. {\it
(b)} As before, but only for the five blazars for which low and
high VHE $\gamma$ flux states have been observed.
\label{fig:comp:compasllum}}}
\end{figure}

\begin{figure}
\center{\includegraphics[width=\linewidth]{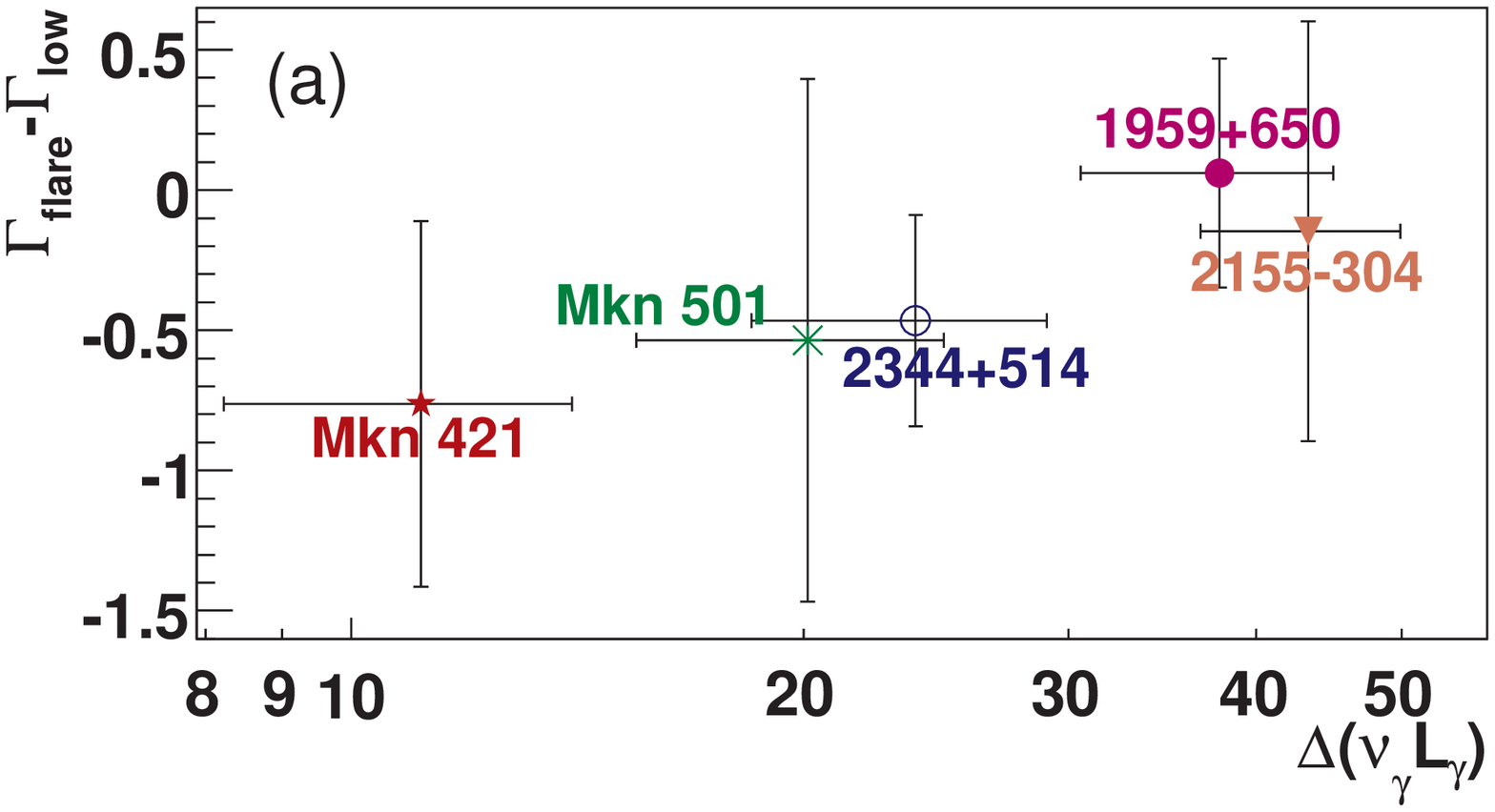}\\
\includegraphics[width=\linewidth]{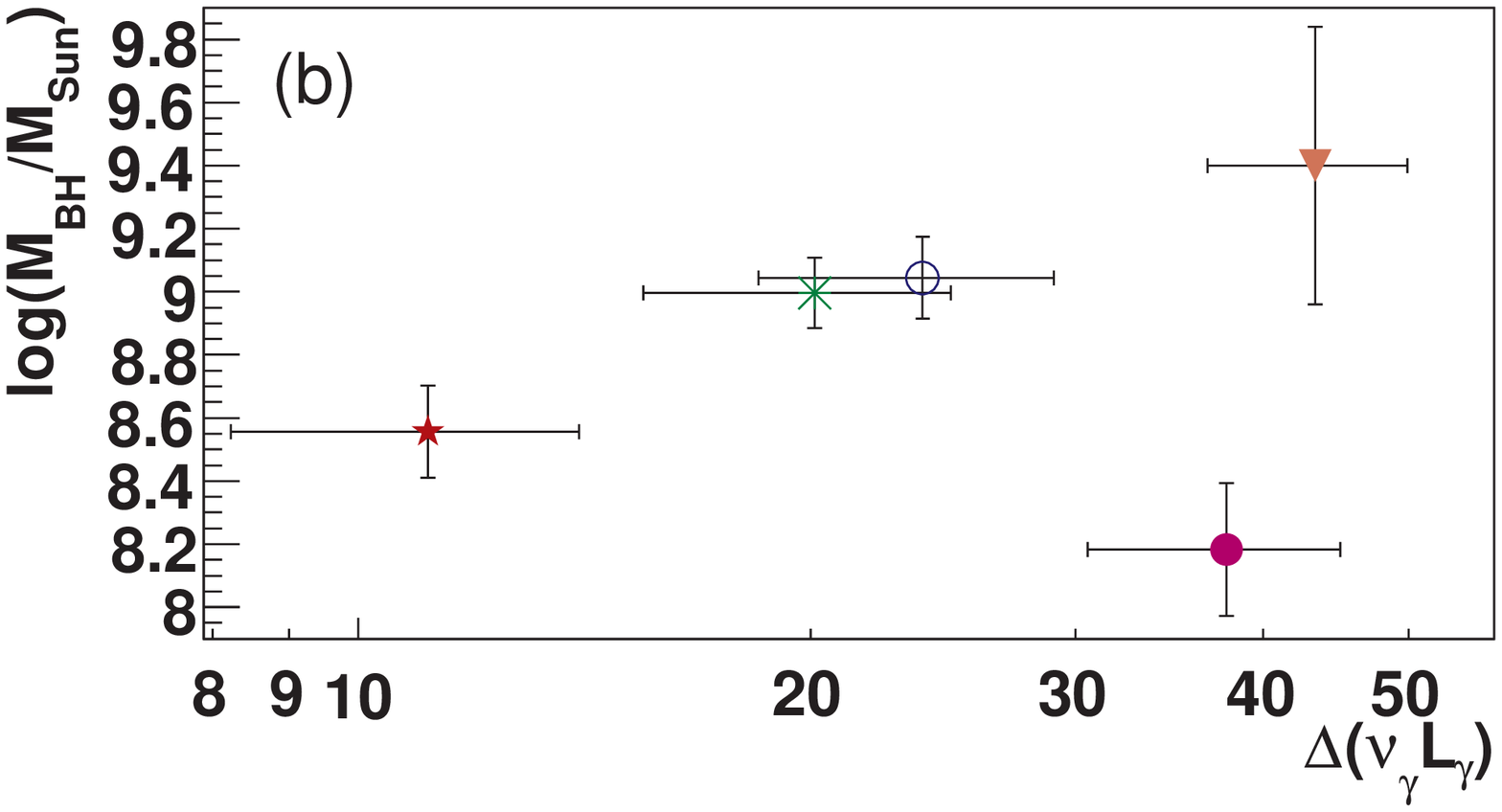}
\caption{{(a)}  Evolution of intrinsic spectral index $\Gamma$
and source luminosity from low to high VHE $\gamma$ flux states:
luminosity ratio $\Delta (\nu_\gamma L_\gamma) = (\nu_\gamma
L_\gamma)_\mathrm{flare}/(\nu_\gamma L_\gamma)_\mathrm{low}$
versus the difference of intrinsic photon indices. {(b)}
luminosity ratio $\Delta (\nu_\gamma L_\gamma) = (\nu_\gamma
L_\gamma)_\mathrm{flare}/(\nu_\gamma L_\gamma)_\mathrm{low}$
versus $M_\bullet$. \label{fig:comp:compasllum2}}}
\end{figure}

Fig.~\ref{fig:comp:compasllum}a relates the intrinsic photon
indices $\Gamma$ to the VHE $\gamma$-ray luminosities $\nu_\gamma
L_\gamma$. For all objects under study (excluding flare states and
the PG\,1553+113 data), the $\Gamma-\nu_\gamma L_\gamma$
correlation reads as $\Gamma = \Gamma_0 + m \cdot \log_{10}
(\nu_\gamma L_\gamma)$ with $\Gamma_0 = 38.5 \pm6.8$ and
$m=-0.82\pm0.15$. The fit which led to this correlation has a
$\chi^2_\mathrm{red} = 40.18/14$. The line at $\Gamma=2$ denotes
the spectral slope at which a maximum in the SED occurs. Looking
at the spread of the observations in $\Gamma$ as a function of
$\nu_\gamma L_\gamma$, I notice that the distribution sharpens
towards $\Gamma=2$. This might reflect that the highest luminosity
occurs at $\Gamma=2$. Thus the spread of the data reflects the
spread of the shapes of the high energy peaks. The general
behaviour above $\Gamma\approx2$ -- the higher the VHE
$\gamma$-ray luminosity, the harder the spectrum -- can within SSC
models be described with a moving IC peak towards higher energies
with increasing luminosity. The five blazars with observed spectra
at quiescent and flare states partially behave in a similar manner
(Fig.~\ref{fig:comp:compasllum}b): Mkn~421
\citep{krawczinski,bla421}, Mkn~501 \citep{MAGIC-501} and
1ES\,2344+514 \citep{MAGIC2344published} also show a spectral
hardening during high-flux states, flares and outbursts.
Fig.~\ref{fig:comp:compasllum2}a shows the corresponding
luminosity differences $\Delta(\nu_\gamma L_\gamma)$ and slope
differences $\Delta \Gamma$. Mkn\,501 and 1ES\,2344+514 show a
similar change in spectral slope and a dynamical range of $\Delta
(\nu_\gamma L_\gamma) \approx 20$. The luminosity increase of
Mkn~421 observed up to now is much lower with $\Delta (\nu_\gamma
L_\gamma) \approx 10$. While the spectral slope of Mkn\,421 also
hardens with increasing luminosity, 1ES\,1959+650 and
PKS\,2155-304 show within errors no variation in their spectral
slopes during flares, while their luminosities increase rather
drastically by a factor of 40 and 50, respectively. When plotting
the luminosity difference versus the BH masses
(Fig.~\ref{fig:comp:compasllum2}b), a trend, although broken by
1ES~1959+650, towards blazars with more massive BHs showing higher
dynamical ranges of their emission, is found.

\subsection{A test for selection effects: redshift dependencies?} \label{sect:intrinsicspectraz}

\begin{figure*}
\center{\includegraphics[width=\linewidth]{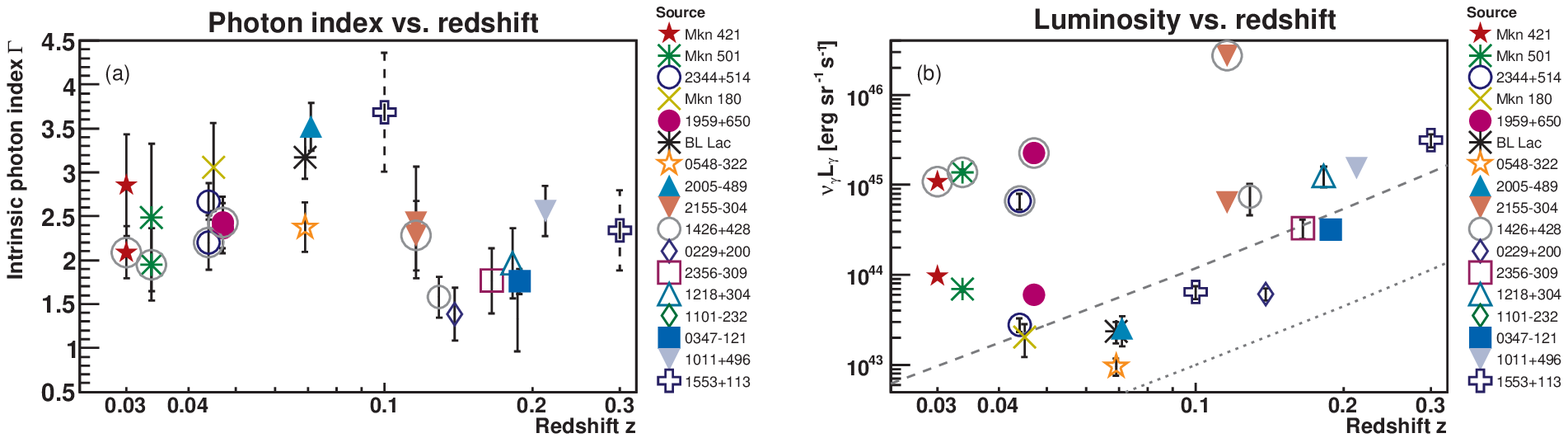}
\caption{Correlations of redshift with {(a)} the intrinsic
photon index and {(b)} VHE $\gamma$-ray luminosity. For
Mkn~421, Mkn~501, 1ES\,2344+514, PKS~2155-304 and 1ES\,1959+650
flare flux levels are also included in the plots; the
corresponding data points are marked by additional grey circles.
The two data points for PG\,1553+113 (open crosses) are for
assumed redshifts of $z=0.1$ and $z=0.3$, respectively. The dotted
curve in the right Figure {(b)} marks the sensitivity
\citep[e.g. ][]{HESS1553} limit of current IACTs (selection
effect);
the dashed curve indicates the sensitivity of previous (before
2002) IACTs. \label{fig:comp:compaspecsz}}}
\end{figure*}

Possible correlations of $\Gamma$ and $\nu_\gamma L_\gamma$ with
the redshift $z$ are not expected, but may identify selection
effects in the data set and/or an inaccurate EBL model.
Conspicuously, very hard ($\Gamma \ll 2.0$) spectra have up to now
been only reconstructed for rather distant ($z\ga 0.1$) blazars
(Fig.~\ref{fig:comp:compaspecsz}a), a trend also observed by
\citet*{sbs07}, who used a smaller set of VHE blazars. At the same
time, none of the measured spectra of nearby sources shows
$\Gamma$ much smaller than 2.0, although for these blazars no
strong EBL modifications apply and the measured spectra should not
differ substantially from the intrinsic ones in the energy region
studied here. Why are only blazars with rather hard intrinsic
spectra visible at large distances ($z>0.1$)? Soft spectra
certainly fall more easily below the current instrumental
sensitivity limits. Another explanation for the prevalent hard
spectra at large $z$ is an overcorrection of the EBL attenuation
effects.

Fig.~\ref{fig:comp:compaspecsz}b shows the corresponding
distribution of VHE luminosities $\nu_\gamma L_\gamma$ as a
function of $z$. Two curves indicate the sensitivity limit of
current IACTs \citep[e.g. ][]{HESS1553}, a significant detection
of 1 per cent of the flux of a Crab nebula-like source in 25
hours, and the sensitivity of previous (before 2002) IACTs of
$\approx 10$ per cent of this flux. Interestingly, most of the
sources found at $z>0.05$ seem to be rather low-luminosity sources
in the sense that their luminosity is not much higher than the
current instrumental sensitivity allows for. This means that not
only the substantially lower energy thresholds of the current
IACTs ($\la 100$~GeV as compared to $\approx 300$~GeV until 2004),
but also their increased sensitivity enabled some of the new
blazar discoveries. Exceptions to these detections close to the
current sensitivity limit are (trivially) the six VHE blazars
discovered before 2002, 1ES\,1011+496 (but discovered during an
optical flare, while its low-flux state seems to lie below the
current instrumental reach, \citealt{1011,MAGIC-UL-Paper}), and
1ES\,1218+304, which in fact seems to have been detected in a
non-flare state. Note further that H\,1426+428 has not yet been
detected after 2002 (e.g. \citealt{veul,MAGIC-UL-Paper}), which
also currently places it below sensitivity limits.

\subsection{Correlations of intrinsic VHE $\gamma$-ray emission parameters with the black
hole mass} \label{sect:intrinsicspectra}

The $\gamma$-ray production is thought to take place at shock
fronts inside the AGN jets at very close (sub-parsec) distances
from the central BH \citep{jester,uchiyama}. While the jet
production and collimation mechanism is still elusive, accepted
models are generally based on magnetohydrodynamic
\citep{1982MNRAS.199..883B,Kudoh} or electromagnetic jet models.
In the latter, a Poynting flux dominated flow is launched from a
Kerr BH \citep{1977MNRAS.179..433B} or from the accretion disc
\citep{1976MNRAS.176..465B}. The conversion from Poynting
dominance into particle dominance is not yet understood. The
properties of the blazar $\gamma$-ray emission are expected to be
connected to the properties of the central BH, like $M_\bullet$
and the BH spin, since scaling laws govern BH physics
\citep{2006Natur.444..730M}, in particular length and time-scales
of flows \citep{1999ARA&A..37..409M,2004inun.conf..175M}, e.g. the
orbital period of the last stable BH orbit. Currently, only
$M_\bullet$ can be reliably estimated; the BH spin remains
inaccessible by large. Moreover, the environment in which the BH
is embedded might be equally important; one of its properties, the
accretion rate, is indirectly accessible through the (radio) jet
power 
\citep{2006ApJ...637..669L}, or from multiwavelength modelling
\citep{mt03}. A previous study of the connection of spectral
properties and $M_\bullet$ for five VHE blazars
\citep{Krawczynski1es1959-2004} did not find any correlations with
the BH mass.

\begin{figure*}
\center{\includegraphics[width=\linewidth]{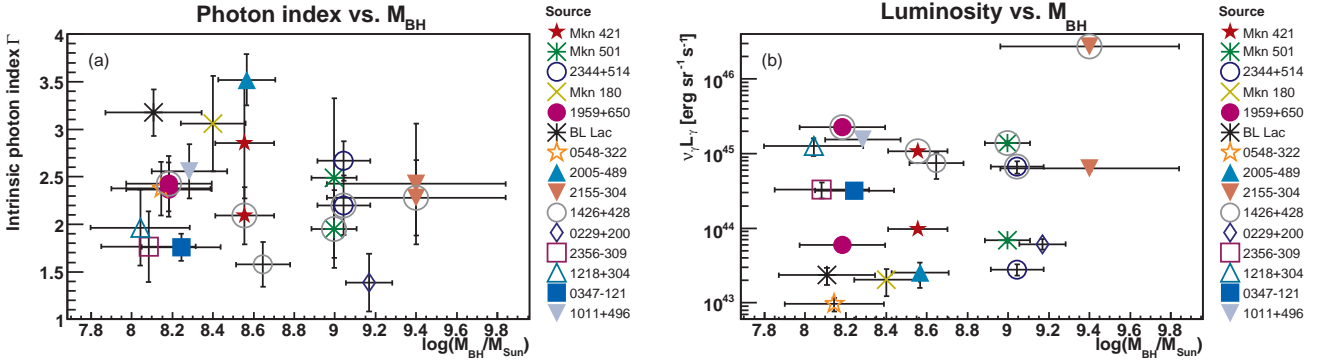}
\caption{Correlations of black hole mass with {(a)} the
intrinsic photon index and {(b)} VHE $\gamma$-ray
luminosity. For Mkn~421, Mkn~501, 1ES\,2344+514 and 1ES\,1959+650
flare flux levels are also included in the plots; the
corresponding data points are marked by additional grey circles.
\label{fig:comp:compaspecs}}}
\end{figure*}

For 15 VHE blazars with known BH masses, I neither find a
correlation between $M_\bullet$ and the spectral slope $\Gamma$
(Fig.~\ref{fig:comp:compaspecs}a), nor between $M_\bullet$ and the
VHE $\gamma$-ray luminosity (Fig.~\ref{fig:comp:compaspecs}b).
While the VHE blazar set disfavours a dependency of the VHE
$\gamma$-ray emission properties studied here on $M_\bullet$, the
uncertainties of the $M_\bullet$ determination are still rather
large and might conceal otherwise interesting physics. In
particular, indirectly inferred $\sigma$ values from the
fundamental plane or bulge luminosity measurements have rather
large uncertainties. To improve on the $M_\bullet$ uncertainties,
it would be desirable to obtain direct measurements of $\sigma$
for all VHE blazars. Moreover, VHE emission properties may also
depend sensitively on the BH spin, the accretion rate, or on
properties of the acceleration region in the jet. Also results on
timing properties (see Sect.~\ref{sect:dcsect}) support such
claims.

\subsection{An upper distance limit for PG 1553+113}
\label{sect:comp:501lumilim}
The redshift determination for blazars is challenging, as these
AGN generally exhibit only weak spectral lines. In several
attempts, no emission or absorption lines could be found in the
optical/IR spectrum of PG\,1553+113; see e.g. \citet{1553icrc}
for results of a recent Very Large Telescope (VLT) campaign.
The frequently cited initial determination of $z=0.36$
\citep{MillerGreen} was found to be based on a misidentified
emission line and could not be reproduced
\citep*{Falomo,1994ApJS...93..125F}. VLT optical spectroscopy
\citep{Sbar} yields a lower limit of $z>0.09$, while the analysis
of Hubble Space Telescope images leads to the prediction of a
redshift in the range of $z=0.3-0.4$ \citep*{treves-07}. Indirect
methods employing the maximum slope
\citep{HESS1553,MAGIC1553published} or the shape
\citep{2007ApJ...655L..13M} of the VHE $\gamma$-ray spectrum find
upper limits of $z<0.74$ and $z<0.42$, respectively.

\begin{figure*}
\center{\includegraphics[width=.66\linewidth]{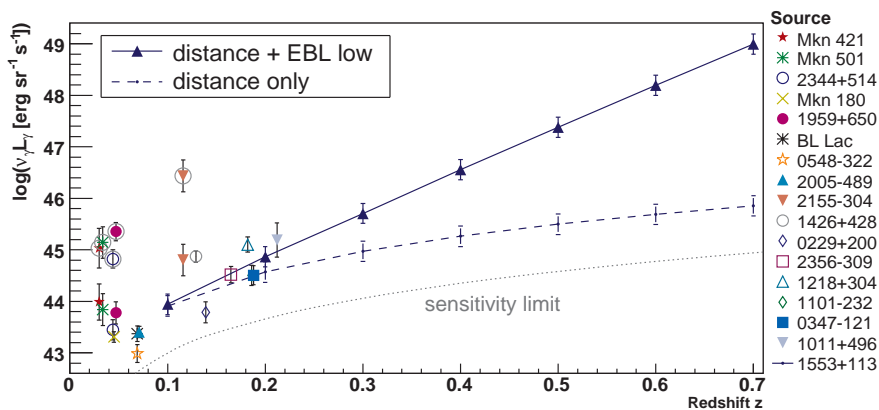}
\caption{Luminosity evolution for PG\,1553+113 assumed at
different distances. Together with the luminosities of the other
known VHE $\gamma$-ray emitting blazars, the luminosity of
PG\,1553+113 as a function of its assumed distance is shown. The
solid curve includes both the distance and the EBL attenuation
effect, the latter calculated using the `low-IR' model given in
\citet{kneiske4}. The dashed curve illustrates how weak the effect
only by increasing the distance is. The dotted curve indicates the
current IACT sensitivity limit neglecting EBL attenuation effects.
Blazars in flaring state, marked with additional gray circles,
were ignored for determining the redshift limit (see text).
 \label{fig:comp:1553lumilim}}}
\end{figure*}
With increasing distance, the luminosity of PG\,1553+113 has to
increase stronger than quadratic due to EBL $\gamma\gamma$
absorption as to sustain the measured VHE flux.
Fig.~\ref{fig:comp:1553lumilim} shows the source luminosity (1)
when only considering the distance effect and (2) when also taking
into account the EBL absorption. Due to the exponential behaviour
of the EBL attenuation, the latter effect is by far dominant. I
assume here that PG\,1553+113 is an `off the shelf' blazar, i.e.
with no extraordinarily high luminosity $\nu_\gamma L_\gamma$.
This assumption is difficult to quantify; given the overall
dynamical range of the (non-flare) blazar luminosities in this
study of $\approx 75$, I consider the case in which the luminosity
of PG\,1553+113 is not more than 30 (1000) times higher than the
highest luminosity found in the sample. Further, PG\,1553+113 has
in two years of observations not shown any apparent flaring
behaviour; within a factor of three, the measured flux was
constant \citep{HESS1553,MAGIC1553published,icrc1553hess}.
Therefore I consider only sources in non-flare states, of which
1ES\,1218+304 with $\nu_\gamma L_\gamma=1.3\times10^{45}\,
\mathrm{erg}\,\mathrm{s}^{-1}\,\mathrm{sr}^{-1}$ is the most
luminous one. A 30-times higher luminosity then implies a limit of
$z<0.45$, while an extreme luminosity of $\nu_\gamma
L_{\gamma}=1.3\times10^{48}\,
\mathrm{erg}\,\mathrm{sr}^{-1}\,\mathrm{s}^{-1}$ yields a limit of
$z<0.64$. These limits do not only depend on a good knowledge of
the EBL over a wide range in redshift, but also on the assumed
maximum VHE blazar luminosity that strongly depends on the Doppler
factor $\delta$. In any case, either a strikingly high luminosity
or a very high $\delta$ is needed to explain the observations
should PG\,1553+113 be more distant than $z\ga0.35$. Note that
$\delta \la 20$ suffices for most of the blazars modelled up to
now during non-outburst times, as also for flare observations
\citep[e.g. Mkn\,421, ][]{Maraschi60}. SSC modelling for
PG\,1553+113 resulted in $\delta=21$
\citep{costamante,MAGIC1553published}.

The difficulties in finding emission and absorption lines might
indicate a very close alignment of the jet axis of PG~1553+113 to
our line of sight. Very-long baseline interferometry (VLBI)
imaging is available for some VHE HBLs detected before 2002
\citep{1999ApJ...525..176P,2002ApJ...579L..67E,2004ApJ...600..115P}
and for BL~Lacertae \citep*{2000ApJS..129...61D}, while additional
measurements of the recently found VHE blazars are underway
\citep*{PinerAK}. VLBI essentially confirms the close alignment of
the jet to our line of sight on sub-parsec scale and finds opening
angles of typically few degrees. These results cannot yet be used
for quantitative correlation studies though.

We have to eventually consider the possibilities that PG\,1553+113
is a rather distant source and the blazar populations at large
distances show significantly different properties than the
close-by objects at $z<0.2$, and that such very extreme objects
are so rare that a sufficiently large volume had to be probed to
find one of them.

\subsection{Correlations with the X-ray duty cycle and the VHE variability time-scale}
\label{sect:dcsect}
\begin{figure*}
\includegraphics[width=.64\linewidth]{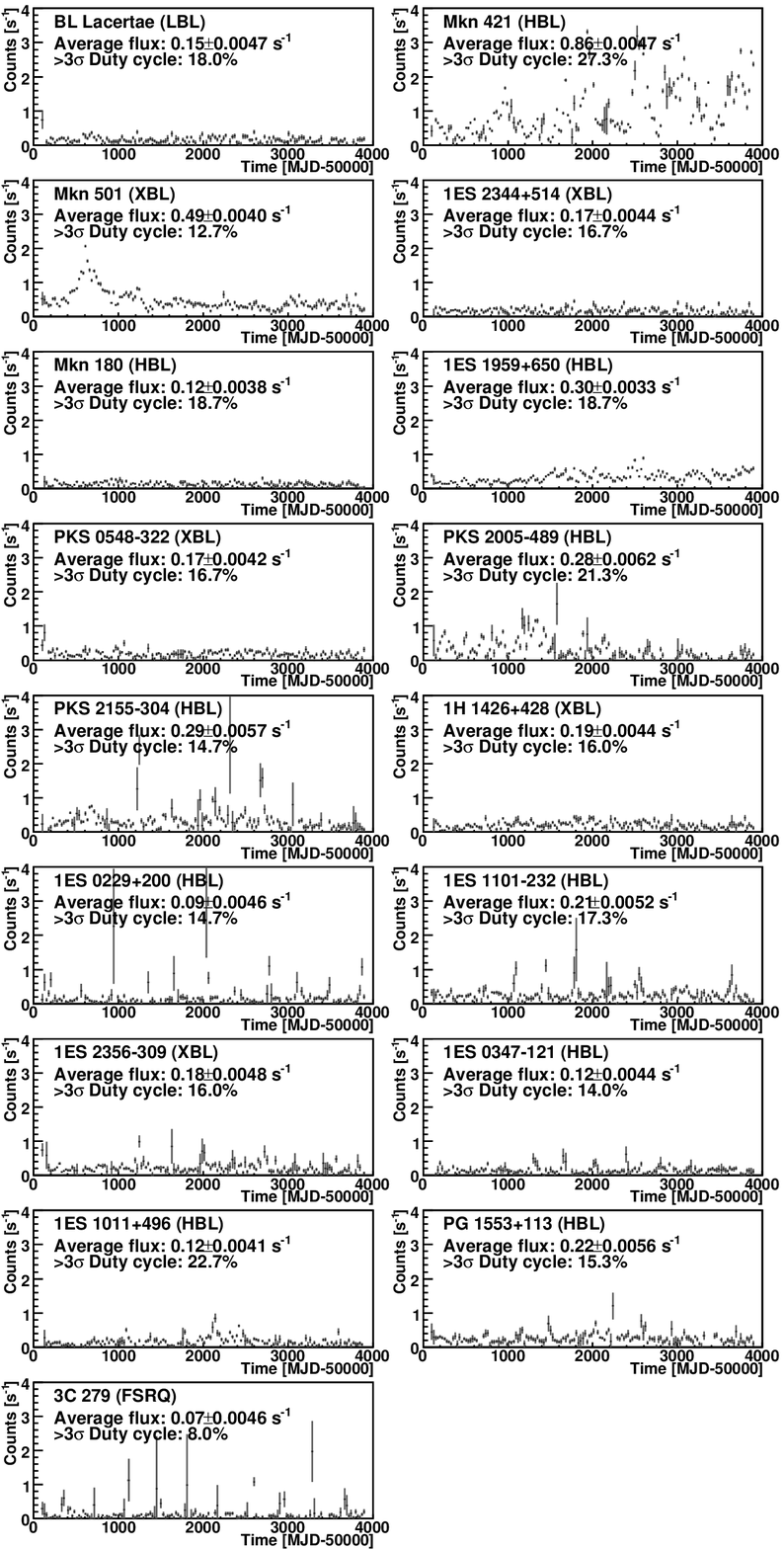}
\caption{X-ray (2-10~keV) light curves of VHE $\gamma$-ray
emitting blazars. The flare duty cycle, i.e. the fraction of time
in which the respective object significantly exceeds its average
X-ray flux by 50 per cent is shown. For convenience the ranges of
the vertical axes are fixed for all plots. FSRQ: Flat spectrum
radio quasar, LBL: low-frequency peaked BL~Lac object; HBL:
high-frequency peaked BL~Lac object, synchrotron peak in the
UV/X-ray range; XBL: extreme BL~Lac object.
\label{fig:comp:compadutycycle}}
\end{figure*}

\begin{figure*}
\center{\includegraphics[width=\linewidth]{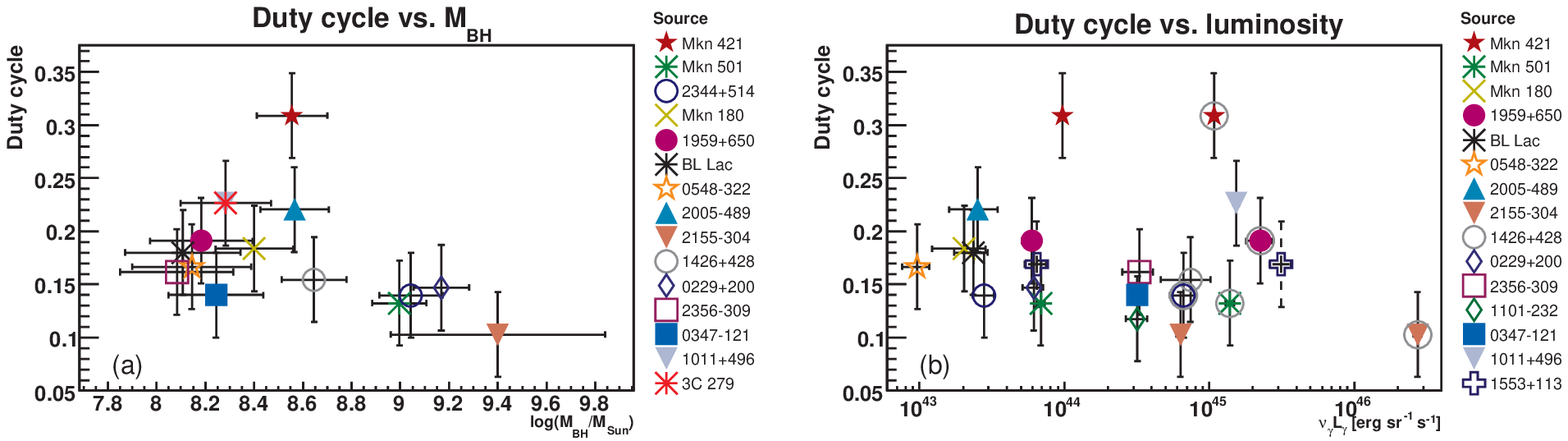}
\caption{Correlations of the X-ray duty cycle with {(a)} the
BH mass and {(b)} the VHE
luminosity.\label{fig:comp:compaspecstime}}}
\end{figure*}

Following a method described in \citet{Krawczynski1es1959-2004}, I
determine the time fraction (`duty cycle') for which the
$(2-10)\,\mathrm{keV}$ X-ray flux exceeds the average flux by 50
per cent. In this paper, for a blazar to be regarded `on duty'
this deviation is additionally required to be significant on the
$3\sigma$ level. 2-10 keV X-ray light curves are obtained from the
All-Sky Monitor detector\footnote{available at
http://xte.mit.edu/} on board the Rossi X-ray Timing Explorer
({\it RXTE}) and are available from 1996 January 5 on.
Fig.~\ref{fig:comp:compadutycycle} shows the corresponding light
curves along with the resulting duty cycles. Objects which are
classified as extreme BL~Lacs (Mkn~501, 1ES\,2344+514,
H\,1426+428, H\,2356-309 and PKS\,0548-322,
\citealt{2001A&A...371..512C}) show substantially higher activity
than the other blazars. Note also the outstanding role of Mkn 421
with its substantially higher duty cycle as compared to all other
objects. Fig.~\ref{fig:comp:compaspecstime}a shows the duty cycle
as a function of the BH mass for the VHE blazars. No trends are
visible, except for the observation that the three blazars with
the most massive BHs ($\log(M_\bullet/\mathrm{M}_{\sun})>8.8$) do
not have duty cycles in excess of 17 per cent. This goes in line
with a speculation of an anticorrelation between the X-ray flare
duty cycle and $M_\bullet$ seen in the five VHE blazars studied in
\citet{Krawczynski1es1959-2004}.
The distribution of the X-ray duty cycle as a function of
luminosity is found to be rather flat
(Fig.~\ref{fig:comp:compaspecstime}b), supporting the claim that
variability on all scales is a defining property of blazars. Note,
however, that Mkn\,501 and PKS\,2155-304, the objects which show
the fastest variability in VHE $\gamma$-rays, are among the
objects with a rather low duty cycle.

Blazars are characterised by a highly variable emission. In
particular the VHE $\gamma$-ray emission is often found to be more
variable than the emission at other wavelenghts. Still,
comprehensive data on the temporal behaviour of VHE blazar
emission has often been collected only for blazars in outburst,
and thus, over short time spans. Although VHE light curves ranging
over ten years' worth of measurements were collected occasionally
\citep[e.g.][]{MAGIC2344published,MAGIC-501,tlucmad}, the sampling
is only sparse, and continuous, unbiased long-term monitoring
campaigns have not been started until 2005
\citep{magicmonitoring,veritasmonitoring,hessmonitoring}.

\begin{figure*}
\center{\includegraphics[width=\linewidth]{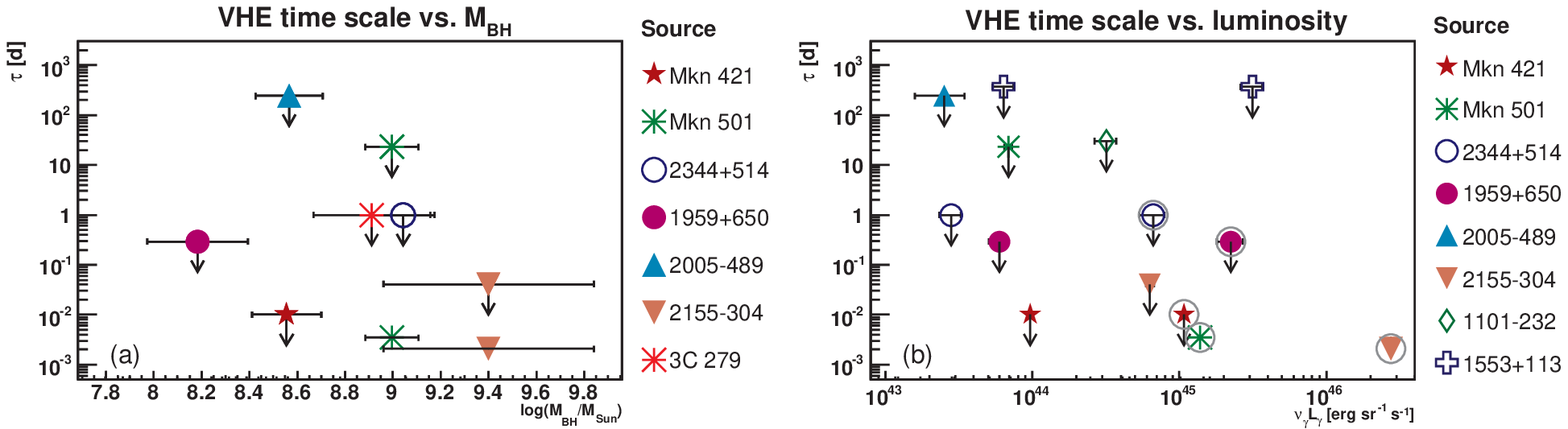}
\caption{Correlations of the VHE $\gamma$ flux doubling time
$\tau$ with {(a)} the BH mass and {(b)} the VHE
luminosity. Upper limits on $\tau$ are taken from
\citet{icrc2005hess} and from the references given in
Tab.~\ref{tab:parvhe}. Additional data points represent the
recently found fast doubling times for Mkn~501 \citep{MAGIC-501} and PKS\,2155-304
\citep{2155flare}. Further for Mkn~421
\citep{Gaidos1996}, 1ES\,2344+514 and 1ES\,1959+650 doubling
times during flare states are also included, which are marked by
additional grey circles. The two data points for PG\,1553+113
represent assumed redshifts of $z=0.1$ (lower luminosity) and
$z=0.3$ (higher luminosity), respectively.
\label{fig:comp:compaspecstime2}}}
\end{figure*}

Variability time-scales $\tau$ are intimately linked to the
extension of the region $R$ from which the observed emission
originates by the causality condition $R \la \delta c \tau /
(1+z)$. \citet{Cui2004} has suggested that the flare hierarchy
seen in Mkn~501 long-term data implies a scale-invariant nature of
the flare process and that there might not be any fundamental
difference among long, intermediate and rapid flares. Although not
thoroughly understood, the flares in blazars might be related to
internal shocks in the jet \citep{ReesM87,Spada} or to major
ejection events of new components of relativistic plasma into the
jet \citep*{BoettcherAA,MasKirk}. Different flare time-scales thus
may be caused by a hierarchy of inhomogeneities in the jet,
energised so as to produce flares. As a timing property of the VHE
$\gamma$-ray emission, the minimum flux doubling times for the VHE
blazars were collected from literature and plotted versus
$M_\bullet$ (Fig.~\ref{fig:comp:compaspecstime2}a) and the VHE
luminosity (Fig.~\ref{fig:comp:compaspecstime2}b).
Because of the very limited data base of VHE variability
measurements, which also is biased towards high-flux states and
outbursts, time-scales are mostly given as upper limits, which
disables strong conclusions.

The observed VHE flux doubling times do not scale with the BH mass
(Fig.~\ref{fig:comp:compaspecstime2}a), which may simply mean that
(1) the flaring mechanism is working in a much smaller region than
the BH radius/least stable orbit and more importantly (2) the BH
and its properties as such do not influence the flaring process
substantially, and the embedding environment of the BH and the jet
environment play more dominant roles (e.g., the accretion power of
the system). The extremely short doubling times of $\tau<5$~min
recently found for Mkn~501 \citep{MAGIC-501} and PKS~2155-304
\citep{2155flare} clearly do not support such a scaling either;
even more, they are vastly incompatible with the rotation period
of particles on the last stable orbit, which e.g. for the BH in
Mkn~501 is $T=8.4$~d. Additionally, the size scale implied by
$\tau<5$~min requires large Doppler factors $\delta$ in the
$\gamma$-ray production regions as to avoid self-absorption. In
contrast to the expected scaling behaviour of the flow properties
around BH with their masses, the three AGN that host rather
massive BHs, PKS\,2155-304, Mkn~501 and Mkn~421 were found to
exhibit the shortest variability time-scales. Until minute-scale
flaring was found in PKS\,2155-304, one could have argued that
short time variability could only be measured for high fluxes due
to the proximity of the respective sources: Mkn 421 and Mkn 501
are the closest blazars at $z<0.034$, while PKS~2155-304 is
located at $z=0.116$. The present observations could, however,
still be the cause of a selection effect: Minute-scale flares were
found only during exceptional high-flux states of the respective
sources so far.

\begin{figure*}
\center{\includegraphics[width=.78\linewidth]{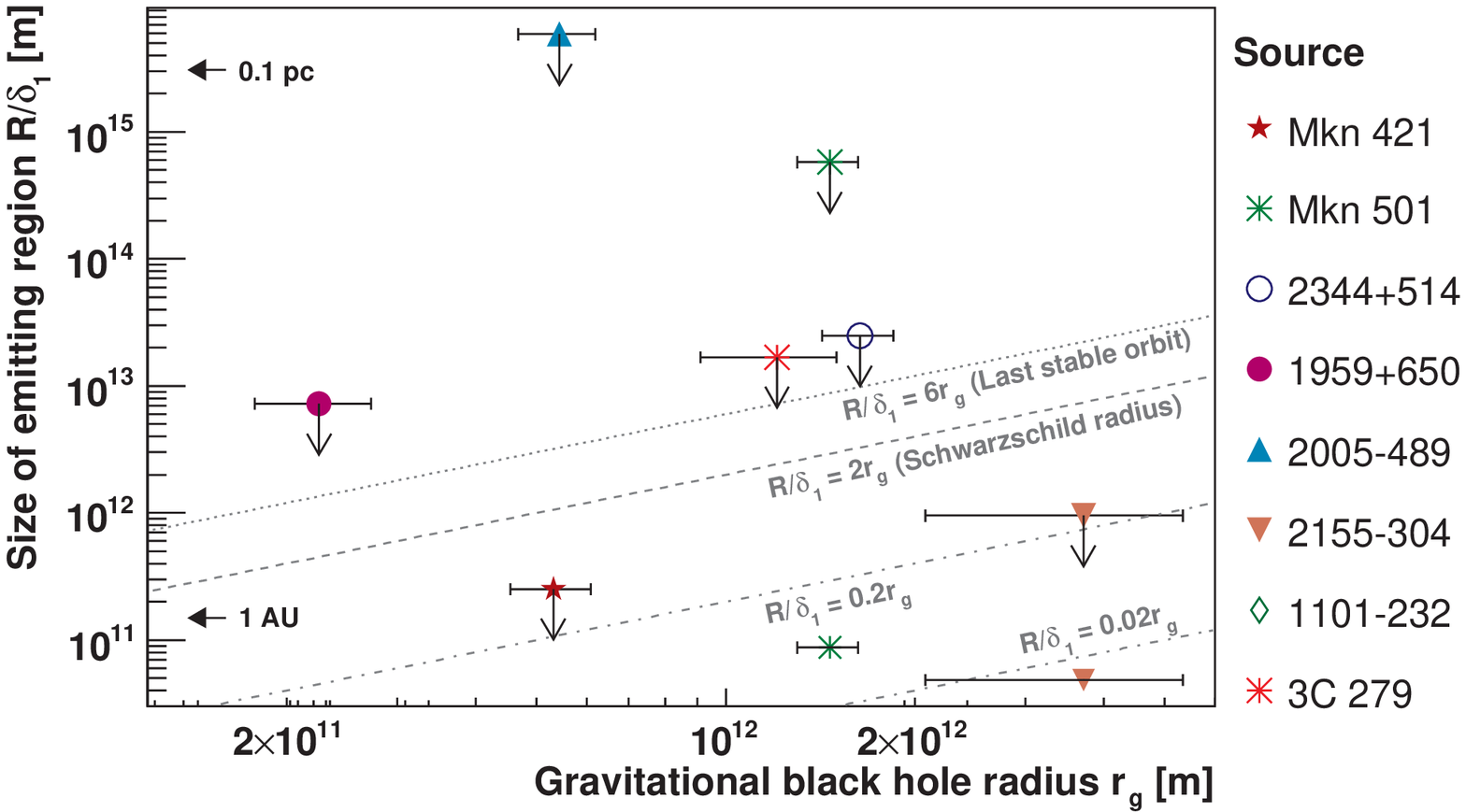}
\caption{Gravitational radius $r_\mathrm{g}$ vs. upper limits on
the size of the VHE $\gamma$-ray emission region $R$ determined by
the causality condition. $\delta_1 = 10 \cdot \delta$ denotes the
Doppler factor. The lines correspond to emission region sizes of
the last stable orbit radius (dotted line), of the Schwarzschild
radius (dashed line), and of 10 per cent and 1 per cent of the
Schwarzschild radius, respectively (dot-dashed lines).
\label{fig:comp:geometrical}}}
\end{figure*}

In Fig.~\ref{fig:comp:geometrical} I translate the BH masses into
the corresponding gravitational radii $r_\mathrm{g}=GM/c^2$
(=$0.5$~Schwarzschild radius) and the observed minimal variability
time-scales into the sizes of the emission regions $R \la \delta c
\tau / (1+z)$, assuming $\delta=10$. For blazars in which fast
variability has been observed, the extreme compactness of the
emission region is apparent, being clearly comparable to or
smaller than the Schwarzschild radius of the central BH. Even if
the flares were driven by extremely large Doppler factors, say
$\delta\approx100$, this would still result in emission regions
smaller than the Schwarzschild radius for the Mkn\,501 and
PKS\,2155-304 flares.

\section{Concluding remarks}
\label{sect:conc}
Before the new generation of IACT became operational, only six
firmly detected extragalactic VHE $\gamma$-ray sources were known
\citep[e.g.,][]{Moriicrc}; not for all of them differential energy
spectra had been inferred. To date, there are 17 BL Lac objects
and the FSRQ 3C~279 known to emit VHE $\gamma$-rays. This
substantially enlarged set  of blazars called for a synoptic
study. I collected and derived intrinsic properties of the VHE
$\gamma$-ray emission (luminosity, spectral hardness, temporal
properties) and further included X-ray, optical and radio
emission properties. As an accessible property of the BHs in the
centres of the blazar host galaxies, the BH mass was also used.
The studies yield the following results.
\begin{itemize}
\item[(i)] So far, only blazars with $M_\bullet \ga 10^8
\mathrm{M}_{\sun}$ show VHE $\gamma$ emission. Whether this
experimental finding constitutes a selection effect or will reveal
interesting physics, remains to be seen.
\item[(ii)] The VHE luminosity of the blazars under study and the
corresponding X-ray luminosity show a hint of a correlation, as
expected from leptonic acceleration/SSC models.
\item[(iii)] A correlation between the spectral slope in the VHE region
and the peak location of the synchrotron peak is found. Such a
correlation is also expected from SSC models.
\item[(iv)] There were no correlations found between the $\gamma$
emission properties and the $M_\bullet$ of the galaxies that host
the blazars. Also, no correlation of $M_\bullet$ could be observed
with the flare duty cycles and the flare time-scales. Thus, VHE
$\gamma$-ray emission properties may not dominantly depend on
$M_\bullet$. Other possibly interesting BH parameters are not yet
within instrumental reach.
\item[(v)] There is an indication that the VHE $\gamma$ luminosity is
correlated with the spectral hardness. This correlation can be
formulated as a decrease of $\Delta \Gamma \approx 0.82$ per
decade of luminosity.
\item[(vi)] This behaviour is also found for some individual blazars
that were observed in different emission states: There are
indications that in some variable sources the observed spectra
become harder with increasing luminosity, while in others no
hardening is found.
\end{itemize}

Investigations of the temporal properties of the X-ray emission
show that the blazars with the most massive BHs have rather low
duty cycles. The temporal behaviour of the VHE $\gamma$-ray
emission is not well studied for most of the VHE blazars, but the
present data do not seem to support a scaling of the flux doubling
time-scales with the BH masses.

Obviously we are still dealing with a low number of sources and
certainly with an incomplete source sample, which leaves regions
in the parameter space empty. Nevertheless, the rather impressive
number of 17 VHE blazars has permitted first comparative studies. 
It also allows
conclusions concerning EBL models: A hint at a marginal
correlation between the intrinsic spectral hardness and the source
distance is likely due to an EBL over-prediction. This result was
also recently quantified by \citet{HESSAGNNature} using the two
distant blazars 1ES\,1101-232 and H~2356-309. In clarifying the
situation, more distant sources particularly with intrinsically
hard spectra (as e.g. 1ES\,0229+200) will play a decisive role;
PG\,1553+113 with its yet undetermined distance might also turn
out to be a good candidate for such analyses once a measurement of
its redshift succeeds. In the meantime, the observed luminosity
distribution of the studied VHE $\gamma$ blazars was used to
constrain the unknown distance of PG\,1553+113 by assuming that the
properties of this blazar are not too different from the most
extreme objects in the blazar sample. Conversely, a large distance
of PG\,1553+113 implies an unusually high luminosity or an
unusually high Doppler factor.

It can be said that the era of blazar astronomy has been entered
-- astronomy being understood as the study of generic properties
of a given class of objects. With the currently known VHE
blazars and the given instrumental sensitivity, VHE blazar
astronomy starts to become less biased: Not only blazars with hard
spectra or during outbursts have been detected, but also low
(quiescent?) emission states in the VHE range are now being
observed and studied.

With continuing discoveries of new sources, it will become easier
to disentangle EBL absorption effects and intrinsic absorption
effects from the measured spectra. Already currently, the VHE
blazar sample contains groups of objects at very similar
distances, e.g. 1ES\,1218+304 and 1ES\,1101-232 ($\Delta
z=0.004$), or the triplet 1ES\,2344+514, Mkn~180 and
1ES\,1959+650 ($\Delta z=0.003$). Such groups are ideal for
direct comparisons between the respective individual objects, as
they are subject to very similar EBL attenuation.

Since 2003, on average three new blazars have been detected per
year. When comparing this rate to the ten years it took to
discover the first six VHE blazars, one can with good reason hope
to accumulate many more sources in the near future, which will
refine synoptic blazar studies as the pioneering one conducted
here. In addition, the first observation of new object classes
like LBLs and FSRQs will certainly help to sharpen the VHE view of
a possible blazar sequence and the underlying physics in the near
future.

\section*{Acknowledgments}
Luigi Costamante kindly provided the full object list from
\citet{costamante}. I would like to thank Eckart Lorenz, Nina
Nowak, W\l odek Bednarek and Hinrich Meyer for fruitful
discussions on this study and the {\it RXTE} team for providing
the all-sky monitor X-ray data. The financial support by Max
Planck Society is gratefully acknowledged. This research has made
use of NASA's Astrophysics Data System.


\begin{thebibliography}{}

\bibitem[\protect\citeauthoryear{Aharonian}{Aharonian}{2000}]{AharonianHad} Aharonian F.~A., 2000, NewA, 5, 377
\bibitem[\protect\citeauthoryear{Aharonian et~al.,}{Aharonian et~al.}{1999a}]{AharonianMkn501-1999a} Aharonian F., et~al. (HEGRA collaboration), 1999a, \aap, 342, 69
\bibitem[\protect\citeauthoryear{Aharonian et~al.,}{Aharonian et~al.}{1999b}]{AharonianMkn4211999} Aharonian F.~A., et~al. (HEGRA collaboration), 1999b, \aap, 350, 757
\bibitem[\protect\citeauthoryear{{Aharonian} et~al.,}{{Aharonian} et~al.}{2001a}]{2001A&A...366...62A} {Aharonian} F.~A., et~al. (HEGRA collaboration), 2001a, \aap, 366, 62
\bibitem[\protect\citeauthoryear{{Aharonian} et~al.,}{{Aharonian} et~al.}{2001b}]{Aharonian2001} {Aharonian} F.~A., et~al. (HEGRA collaboration), 2001b, \apj, 546, 898
\bibitem[\protect\citeauthoryear{{Aharonian} et~al.,}{{Aharonian} et~al.}{2002}]{2002AA...384L..23A} {Aharonian} F., et~al. (HEGRA collaboration), 2002, \aap, 384, L23
\bibitem[\protect\citeauthoryear{Aharonian et~al.,}{Aharonian et~al.}{2004}]{Tluczykont} Aharonian F.~A., et~al. (HEGRA collaboration), 2004, \aap, 421, 529
\bibitem[\protect\citeauthoryear{Aharonian et~al.,}{Aharonian et~al.}{2005a}]{Aharonian2005} Aharonian F.~A., et~al. (H.E.S.S. collaboration), 2005a, \aap, 430, 865
\bibitem[\protect\citeauthoryear{Aharonian et~al.,}{Aharonian et~al.}{2005b}]{HESS2005} Aharonian F.~A., et~al. (H.E.S.S. collaboration), 2005b, \aap, 436, L17
\bibitem[\protect\citeauthoryear{Aharonian et~al.,}{Aharonian et~al.}{2005c}]{2005A&A...441..465A} Aharonian F., et~al. (H.E.S.S. collaboration), 2005c, \aap, 441, 465
\bibitem[\protect\citeauthoryear{Aharonian et~al.,}{Aharonian et~al.}{2006a}]{HESSAGNNature} Aharonian F.~A., et~al. (H.E.S.S. collaboration), 2006a, \nat, 440, 1018
\bibitem[\protect\citeauthoryear{Aharonian et~al.,}{Aharonian et~al.}{2006b}]{HESS1553} Aharonian F.~A., et~al. (H.E.S.S. collaboration), 2006b, \aap, 448, L19
\bibitem[\protect\citeauthoryear{{Aharonian} et~al.,}{{Aharonian} et~al.}{2006c}]{hess87n} {Aharonian} F., et~al. (H.E.S.S. collaboration), 2006c, \sci, 314, 1425
\bibitem[\protect\citeauthoryear{Aharonian et~al.,}{Aharonian et~al.}{2007a}]{1101} Aharonian F., et~al. (H.E.S.S. collaboration), 2007a, \aap, 470, 475
\bibitem[\protect\citeauthoryear{Aharonian et~al.,}{Aharonian et~al.}{2007b}]{2155flare} Aharonian F., et~al. (H.E.S.S. collaboration), 2007b, \apj, 664, L71
\bibitem[\protect\citeauthoryear{Aharonian et~al.,}{Aharonian et~al.}{2007c}]{0347} Aharonian F., et~al. (H.E.S.S. collaboration), 2007c, \aap, 473, L25
\bibitem[\protect\citeauthoryear{Albert et~al.,}{Albert et~al.}{2006a}]{NadiaPaper} Albert J., et~al. (MAGIC collaboration), 2006a, \apj, 639, 761
\bibitem[\protect\citeauthoryear{Albert et~al.,}{Albert et~al.}{2006b}]{MAGIC1218} Albert J., et~al. (MAGIC collaboration), 2006b, \apj, 642, L119
\bibitem[\protect\citeauthoryear{Albert et~al.,}{Albert et~al.}{2006c}]{MAGIC180} Albert J., et~al. (MAGIC collaboration), 2006c, \apj, 648, L105
\bibitem[\protect\citeauthoryear{Albert et~al.,}{Albert et~al.}{2007a}]{MAGIC1553published} Albert J., et~al. (MAGIC collaboration), 2007a, \apj, 654, L119
\bibitem[\protect\citeauthoryear{Albert et~al.,}{Albert et~al.}{2007b}]{MAGIC2344published} Albert J., et~al. (MAGIC collaboration), 2007b, \apj, 662, 892
\bibitem[\protect\citeauthoryear{Albert et~al.,}{Albert et~al.}{2007c}]{MAGIC421published} Albert J., et~al. (MAGIC collaboration), 2007c, \apj, 663, 125
\bibitem[\protect\citeauthoryear{Albert et~al.,}{Albert et~al.}{2007d}]{MAGICBLLac} Albert J., et~al. (MAGIC collaboration), 2007d, \apj, 666, L17
\bibitem[\protect\citeauthoryear{Albert et~al.,}{Albert et~al.}{2007e}]{1011} Albert J., et~al. (MAGIC collaboration), 2007e, \apj, 667, L21
\bibitem[\protect\citeauthoryear{Albert et~al.,}{Albert et~al.}{2007f}]{MAGIC-501} Albert J., et~al. (MAGIC collaboration), 2007f, \apj, 669, 862
\bibitem[\protect\citeauthoryear{Albert et~al.,}{Albert et~al.}{2008}]{MAGIC-UL-Paper} Albert J., et~al. (MAGIC collaboration), 2008, \apj, preprint (arXiv:0706.4453)
\bibitem[\protect\citeauthoryear{Baixeras et~al.,}{Baixeras et~al.}{2004}]{Baixeras} Baixeras C., et~al. (MAGIC collaboration), 2004, Nucl. Instr. Meth. A, 518, 188
\bibitem[\protect\citeauthoryear{Barth, Ho \& Sargent}{Barth et~al.}{2003}]{Barth2003} Barth A.~J., Ho L.~C., Sargent W. L.~W., 2003, \apj, 583, 134
\bibitem[\protect\citeauthoryear{{Bednarek}}{{Bednarek}}{1993}]{1993ApJ...402L..29B} {Bednarek} W., 1993, \apj, 402, L29
\bibitem[\protect\citeauthoryear{Benbow \& B\"uhler}{Benbow \& B\"uhler}{2007}]{HESS-UL-ICRC} Benbow W., B\"uhler R. (H.E.S.S. collaboration), 2007, in {Proceedings of the 30th International Cosmic Ray Conference}, Merida, preprint (arXiv:0709.4598)
\bibitem[\protect\citeauthoryear{Benbow, Boisson, B\"uhler \& Sol}{Benbow et~al.}{2007}]{icrc1553hess} Benbow W., Boisson C., B\"uhler R., Sol H. (H.E.S.S. collaboration), 2007, in {Proceedings of the 30th International Cosmic Ray Conference}, Merida, preprint (arXiv:0709.4602)
\bibitem[\protect\citeauthoryear{Bender \& Kormendy}{Bender \& Kormendy}{2003}]{Bender-03} Bender R., Kormendy J., 2003, in Shaver P., Dilella L., Gim\'{e}nez A., eds, Astronomy, Cosmology and Fundamental Physics, p.~262 (Berlin and Heidelberg: Springer)
\bibitem[\protect\citeauthoryear{{Bettoni}, {Falomo}, {Fasano}, {Govoni}, {Salvo} \& {Scarpa}}{{Bettoni} et~al.}{2001}]{Bettoni} {Bettoni} D., {Falomo} R., {Fasano} G., {Govoni} F., {Salvo} M., {Scarpa} R., 2001, \aap, 380, 471
\bibitem[\protect\citeauthoryear{{Blandford}}{{Blandford}}{1976}]{1976MNRAS.176..465B} {Blandford} R.~D., 1976, \mnras, 176, 465
\bibitem[\protect\citeauthoryear{{Blandford} \& {K\"{o}nigl}}{{Blandford} \& {K\"{o}nigl}}{1979}]{1979ApJ...232...34B}{Blandford} R.~D., {K\"{o}nigl} A., 1979, \apj, 232, 34
\bibitem[\protect\citeauthoryear{{Blandford} \& {Payne}}{{Blandford} \& {Payne}}{1982}]{1982MNRAS.199..883B}{Blandford} R.~D., {Payne} D.~G., 1982, \mnras, 199, 883
\bibitem[\protect\citeauthoryear{{Blandford} \& {Znajek}}{{Blandford} \& {Znajek}}{1977}]{1977MNRAS.179..433B}{Blandford} R.~D., {Znajek} R.~L., 1977, \mnras, 179, 433
\bibitem[\protect\citeauthoryear{{B{\l}a{\.z}ejowski}, {Blaylock}, {Bond} et~al.,}{{B{\l}a{\.z}ejowski} et~al.}{2005}]{bla421}{B{\l}a{\.z}ejowski} M., {Blaylock} G., {Bond} I.~H., et~al., 2005, \apj, 630, 130
\bibitem[\protect\citeauthoryear{{B{\"o}ttcher}, {Mause} \& {Schlickeiser}}{{B{\"o}ttcher} et~al.}{1997}]{BoettcherAA} {B{\"o}ttcher} M., {Mause} H., {Schlickeiser} R., 1997, \aap, 324, 395
\bibitem[\protect\citeauthoryear{{Buckley} et~al.,}{{Buckley} et~al.}{1996}]{1996ApJ...472L...9B} {Buckley} J.~H., et~al. (Whipple collaboration), 1996, \apj, 472, L9
\bibitem[\protect\citeauthoryear{{Catanese} et~al.,}{{Catanese} et~al.}{1998}]{Catanese1998} {Catanese} M., et~al. (Whipple collaboration), 1998, \apj, 501, 616
\bibitem[\protect\citeauthoryear{Chadwick et~al.,}{Chadwick et~al.}{1999}]{Chadwick} Chadwick P.~M., et~al., 1999, \apj, 513, 161
\bibitem[\protect\citeauthoryear{Cogan}{Cogan}{2007}]{VERITAS-UL-ICRC} Cogan P. (VERITAS collaboration), 2007, in {Proceedings of the 30th International Cosmic Ray Conference}, Merida, preprint (arXiv:0709.3695)
\bibitem[\protect\citeauthoryear{Coppi}{Coppi}{1992}]{coppi2} Coppi P.~S., 1992, MNRAS, 258, 657
\bibitem[\protect\citeauthoryear{Cortina et~al.,}{Cortina et~al.}{2005}]{Cortina} Cortina J., et~al. (MAGIC collaboration), 2005, in {Proceedings of the 29th International Cosmic Ray Conference} Vol.~5, Pune, pp 359--362
\bibitem[\protect\citeauthoryear{Costamante \& Ghisellini}{Costamante \& Ghisellini}{2002}]{costamante} Costamante L., Ghisellini G., 2002, \aap, 384, 56
\bibitem[\protect\citeauthoryear{{Costamante}, {Ghisellini}, {Giommi}, {Tagliaferri}, {Celotti}, {Chiaberge}, {Fossati}, {Maraschi}, {Tavecchio}, {Treves} \& {Wolter}}{{Costamante} et~al.}{2001}]{2001A&A...371..512C} {Costamante} L., {Ghisellini} G., {Giommi} P., {Tagliaferri} G., {Celotti} A., {Chiaberge} M., {Fossati} G., {Maraschi} L., {Tavecchio} F., {Treves} A., {Wolter} A., 2001, \aap, 371, 512
\bibitem[\protect\citeauthoryear{Costamante, Benbow, Boisson, Pita \& Sol}{Costamante et~al.}{2007}]{icrc2005hess} Costamante L., Benbow W., Boisson C., Pita S., Sol H. (H.E.S.S. collaboration), 2007, in {Proceedings of the 30th International Cosmic Ray Conference}, Merida, preprint (arXiv:0710.4057, p.~122)
\bibitem[\protect\citeauthoryear{Cui}{Cui}{2004}]{Cui2004} Cui W., 2004, \apj, 605, 662
\bibitem[\protect\citeauthoryear{{Daniel}, {Badran}, {Bond}, {Boyle} et~al.,}{{Daniel} et~al.}{2005}]{2005ApJ...621..181D} {Daniel} M.~K., et~al. (Whipple collaboration), 2005, \apj, 621, 181
\bibitem[\protect\citeauthoryear{{Dar} \& {Laor}}{{Dar} \& {Laor}}{1997}]{1997ApJ...478L...5D} {Dar} A., {Laor} A., 1997, \apjl, 478, L5
\bibitem[\protect\citeauthoryear{{Denn}, {Mutel} \& {Marscher}}{{Denn} et~al.}{2000}]{2000ApJS..129...61D} {Denn} G.~R., {Mutel} R.~L., {Marscher} A.~P., 2000, \apjs, 129, 61
\bibitem[\protect\citeauthoryear{{Dermer} \& {Schlickeiser}}{{Dermer} \& {Schlickeiser}}{1994}]{1994ApJS...90..945D} {Dermer} C.~D., {Schlickeiser} R., 1994, \apjs, 90, 945
\bibitem[\protect\citeauthoryear{{Djorgovski} \& {Davis}}{{Djorgovski} \& {Davis}}{1987}]{1987ApJ...313...59D} {Djorgovski} S., {Davis} M., 1987, \apj, 313, 59
\bibitem[\protect\citeauthoryear{{Donato}, {Ghisellini}, {Tagliaferri} \& {Fossati}}{{Donato} et~al.}{2001}]{2001AA...375..739D} {Donato} D., {Ghisellini} G., {Tagliaferri} G., {Fossati} G., 2001, \aap, 375, 739
\bibitem[\protect\citeauthoryear{{Edwards} \& {Piner}}{{Edwards} \& {Piner}}{2002}]{2002ApJ...579L..67E} {Edwards} P.~G., {Piner} B.~G., 2002, \apjl, 579, L67
\bibitem[\protect\citeauthoryear{{Falomo}}{{Falomo}}{1996}]{1996MNRAS.283..241F} {Falomo} R., 1996, \mnras, 283, 241
\bibitem[\protect\citeauthoryear{Falomo \& Treves}{Falomo \& Treves}{1990}]{Falomo} Falomo R., Treves A., 1990, PASP, 102, 1120
\bibitem[\protect\citeauthoryear{{Falomo}, {Kotilainen} \& {Treves}}{{Falomo} et~al.}{2002}]{2002ApJ...569L..35F} {Falomo} R., {Kotilainen} J.~K., {Treves} A., 2002, \apj, 569, L35
\bibitem[\protect\citeauthoryear{{Falomo}, {Scarpa} \& {Bersanelli}}{{Falomo} et~al.}{1994}]{1994ApJS...93..125F} {Falomo} R., {Scarpa} R., {Bersanelli} M., 1994, \apjs, 93, 125
\bibitem[\protect\citeauthoryear{Fazio \& Stecker}{Fazio \& Stecker}{1970}]{fs70} Fazio G.~G., Stecker F.~W., 1970, Nat, 226, 135
\bibitem[\protect\citeauthoryear{{Ferrarese} \& {Merritt}}{{Ferrarese} \& {Merritt}}{2000}]{2000ApJ...539L...9F} {Ferrarese} L., {Merritt} D., 2000, \apj, 539, L9
\bibitem[\protect\citeauthoryear{{Fossati}, {Maraschi}, {Celotti}, {Comastri} \& {Ghisellini}}{{Fossati} et~al.}{1998}]{fossati} {Fossati} G., {Maraschi} L., {Celotti} A., {Comastri} A., {Ghisellini} G., 1998, MNRAS, 299, 433
\bibitem[\protect\citeauthoryear{Gaidos et~al.,}{Gaidos et~al.}{1996}]{Gaidos1996} Gaidos J.~A., et~al. (Whipple collaboration), 1996, Nature, 383, 319
\bibitem[\protect\citeauthoryear{{Gebhardt}, {Bender}, {Bower}, {Dressler}, {Faber} et~al.,}{{Gebhardt} et~al.}{2000}]{Gebhardt} {Gebhardt} K., {Bender} R., {Bower} G., {Dressler} A., {Faber} S.~M., et~al., 2000, \apj, 539, L13
\bibitem[\protect\citeauthoryear{Goebel, Backes, Bretz, Hayashida, Hsu, Mannheim, Mazin \& Wagner}{Goebel et~al.}{2007}]{magicmonitoring} Goebel F., Backes M., Bretz T., Hayashida M., Hsu C.-C., Mannheim K., Mazin D., Wagner R.~M., 2007, in {Proceedings of the 30th International Cosmic Ray Conference}, Merida, preprint (arXiv:0709.2032)
\bibitem[\protect\citeauthoryear{Gould \& Schr\'{e}der}{Gould \& Schr\'{e}der}{1966}]{gould} Gould R.~J., Schr\'{e}der G., 1966, \prl, 16, 252
\bibitem[\protect\citeauthoryear{{Graham}}{{Graham}}{2007}]{2007MNRAS.379..711G} {Graham} A.~W., 2007, \mnras, 379, 711
\bibitem[\protect\citeauthoryear{{Gu}, {Cao} \& {Jiang}}{{Gu} et~al.}{2001}]{2001MNRAS.327.1111G} {Gu} M., {Cao} X., {Jiang} D.~R., 2001, \mnras, 327, 1111
\bibitem[\protect\citeauthoryear{Hartman, Bertsch, Bloom et~al.,}{Hartman et~al.}{1999}]{hartman} Hartman R.~C., Bertsch D.~L., Bloom S.~D., et~al., 1999, \apjs, 123, 79
\bibitem[\protect\citeauthoryear{Hauser \& Dwek}{Hauser \& Dwek}{2001}]{hauser} Hauser G.~H., Dwek E., 2001, ARA\&A, 39, 249
\bibitem[\protect\citeauthoryear{Hinton}{Hinton}{2004}]{HESStech} Hinton J., 2004, NewAR, 48, 331
\bibitem[\protect\citeauthoryear{{Hogg}, {Baldry}, {Blanton} \& {Eisenstein}}{{Hogg} et~al.}{2002}]{hogg-02} {Hogg} D.~W., {Baldry} I.~K., {Blanton} M.~R., {Eisenstein} D.~J., 2002, arXiv:astro-ph/0210394
\bibitem[\protect\citeauthoryear{{Horan} \& {Finley}}{{Horan} \& {Finley}}{2001}]{2001ICRC....7.2622H} {Horan} D., {Finley} J.~P., 2001, in Proceedings of the 27th International Cosmic Ray Conference, Hamburg, p.~2622
\bibitem[\protect\citeauthoryear{Horan et~al.,}{Horan et~al.}{2002}]{Horan} Horan D., et~al. (Whipple collaboration), 2002, \apj, 571, 753
\bibitem[\protect\citeauthoryear{Horan et~al.,}{Horan et~al.}{2004}]{Horan04} Horan D., et~al. (Whipple collaboration), 2004, \apj, 603, 51
\bibitem[\protect\citeauthoryear{Jester, Harris, Marshall \& Meisenheimer}{Jester et~al.}{2006}]{jester} Jester S., Harris D.~E., Marshall H.~L., Meisenheimer K., 2006, \apj, 648, 900
\bibitem[\protect\citeauthoryear{Kashlinsky}{Kashlinsky}{2005}]{2005PhR...409..361K} Kashlinsky A., 2005, {Phys. Rep.}, 409, 361
\bibitem[\protect\citeauthoryear{Katarzy{\'n}ski, Ghisellini, Tavecchio, Gracia \& Maraschi}{Katarzy{\'n}ski et~al.}{2006a}]{SS3} Katarzy{\'n}ski K., Ghisellini G., Tavecchio F., Gracia J., Maraschi L., 2006a, \mnras, 368, L52
\bibitem[\protect\citeauthoryear{Katarzy{\'n}ski, Ghisellini, Mastichiadis, Tavecchio \& Maraschi}{Katarzy{\'n}ski et~al.}{2006b}]{SS4}Katarzy{\'n}ski K., Ghisellini G., Mastichiadis A., Tavecchio F., Maraschi L., 2006b, \aap, 453, 47
\bibitem[\protect\citeauthoryear{{Kerrick} et~al.,}{{Kerrick} et~al.}{1995}]{1995ApJ...452..588K} {Kerrick} A.~D., et~al. (Whipple collaboration), 1995, \apj, 452, 588
\bibitem[\protect\citeauthoryear{Kneiske, Bretz, Mannheim \& Hartmann}{Kneiske et~al.}{2004}]{kneiske4}Kneiske T.~M., Bretz T., Mannheim K., Hartmann D.~H., 2004, \aap, 413, 807
\bibitem[\protect\citeauthoryear{{Kormendy} \& {Richstone}}{{Kormendy} \& {Richstone}}{1995}]{1995ARAA..33..581K}{Kormendy} J., {Richstone} D., 1995, \araa, 33, 581
\bibitem[\protect\citeauthoryear{Krawczynski}{Krawczynski}{2007}]{veul} Krawczynski H. (VERITAS collaboration), 2007, in {Proceedings of the 30th International Cosmic Ray Conference}, Merida, preprint (arXiv:0710.0089)
\bibitem[\protect\citeauthoryear{Krawczynski et~al.,}{Krawczynski et~al.}{2001}]{krawczinski}Krawczynski H., et~al., 2001, \apj, 559, 187
\bibitem[\protect\citeauthoryear{Krawczynski et~al.}{Krawczynski et~al.}{2004}]{Krawczynski1es1959-2004}Krawczynski H., et~al., 2004, \apj, 601, 151
\bibitem[\protect\citeauthoryear{{Krennrich} et~al.,}{{Krennrich} et~al.}{2001}]{2001ApJ...560L..45K} {Krennrich} F., et~al. (Whipple collaboration), 2001, \apjl, 560, L45
\bibitem[\protect\citeauthoryear{Krennrich et~al.,}{Krennrich et~al.}{2002}]{Krennrich2002}Krennrich F., et~al. (Whipple collaboration), 2002, \apj, 575, L9
\bibitem[\protect\citeauthoryear{Kudoh, Aoki, Koide \& Shibata}{Kudoh et~al.}{1999}]{Kudoh} Kudoh T., Aoki S., Koide S., Shibata K., 1999, Astron. Nachr., 320, 311
\bibitem[\protect\citeauthoryear{Laor}{{Laor}}{2000}]{laor2000} {Laor} A., 2000, \apj, 543, L111
\bibitem[\protect\citeauthoryear{{Liu}, {Jiang} \& {Gu}}{{Liu} et~al.}{2006}]{2006ApJ...637..669L} {Liu} Y., {Jiang} D.~R., {Gu} M.~F., 2006, \apj, 637, 669
\bibitem[\protect\citeauthoryear{{Lynden-Bell}}{{Lynden-Bell}}{1969}]{1969Natur.223..690L} {Lynden-Bell} D., 1969, Nature, 223, 690
\bibitem[\protect\citeauthoryear{{Malkov} \& {Drury}}{{Malkov} \& {Drury}}{2001}]{2001RPPh...64..429M}{Malkov} M.~A., {Drury} L.~O., 2001, Rep. Prog. Phys., 64, 429
\bibitem[\protect\citeauthoryear{Mannheim}{Mannheim}{1993}]{MannheimProton}Mannheim K., 1993, \aap, 269, 67
\bibitem[\protect\citeauthoryear{Maraschi, Ghisellini \& Celotti}{Maraschi et~al.}{1992}]{maraschi}Maraschi L., Ghisellini G., Celotti A., 1992, \apjl, 397, L5
\bibitem[\protect\citeauthoryear{Maraschi \& Tavecchio}{Maraschi \& Tavecchio}{2003}]{mt03}Maraschi L., Tavecchio, F., 2003, \apj, 593, 667
\bibitem[\protect\citeauthoryear{Maraschi et~al.,}{Maraschi et~al.}{1999}]{Maraschi60}Maraschi L., et~al., 1999, \apjl, 526, L81
\bibitem[\protect\citeauthoryear{Mastichiadis \& Kirk}{Mastichiadis \& Kirk}{1997}]{MasKirk} Mastichiadis A., Kirk J.~G., 1997, \aap, 320, 19
\bibitem[\protect\citeauthoryear{Mazin}{Mazin}{2003}]{DanielDiplom}Mazin D., 2003, {Diplomarbeit}, Universit\"at Hamburg
\bibitem[\protect\citeauthoryear{{Mazin} \& {Goebel}}{{Mazin} \& {Goebel}}{2007}]{2007ApJ...655L..13M}{Mazin} D., {Goebel} F., 2007, \apj, 655, L13
\bibitem[\protect\citeauthoryear{{McHardy}, {Koerding}, {Knigge}, {Uttley} \& {Fender}}{{McHardy} et~al.}{2006}]{2006Natur.444..730M} {McHardy} I.~M., {Koerding} E., {Knigge} C., {Uttley} P., {Fender} R.~P., 2006, \nat, 444, 730
\bibitem[\protect\citeauthoryear{{McLure} \& {Dunlop}}{{McLure} \& {Dunlop}}{2002}]{McLure}{McLure} R.~J., {Dunlop} J.~S., 2002, \mnras, 331, 795
\bibitem[\protect\citeauthoryear{Melia \& K\"onigl}{Melia \& K\"onigl}{1989}]{mk89}Melia F., K\"onigl A., 1989, \apj, 340, 162
\bibitem[\protect\citeauthoryear{Metcalf \& Magliocchetti}{Metcalf \& Magliocchetti}{2006}]{mm06}Metcalf R.~B., Magliocchetti M., 2006, \mnras, 365, 101
\bibitem[\protect\citeauthoryear{Miller \& Green}{Miller \& Green}{1983}]{MillerGreen}Miller H.~R., Green R.~F., 1983, \baas, 15, 957
\bibitem[\protect\citeauthoryear{{Mirabel}}{{Mirabel}}{2004}]{2004inun.conf..175M} {Mirabel} I.~F., 2004, in {Sch\"onfelder} V., {Lichti} G., {Winkler} C., eds, ESA SP-552: 5th INTEGRAL Workshop on the INTEGRAL Universe, p.~175 (Noordwijk: ESA)
\bibitem[\protect\citeauthoryear{{Mirabel} \& {Rodr{\'{\i}}guez}}{{Mirabel} \& {Rodr{\'{\i}}guez}}{1999}]{1999ARA&A..37..409M}{Mirabel} I.~F., {Rodr{\'{\i}}guez} L.~F., 1999, ARA\&A, 37, 409
\bibitem[\protect\citeauthoryear{Mori}{Mori}{2003}]{Moriicrc}Mori M., 2003, in {Proceedings of the 28th International Cosmic Ray Conference}, Tsukuba, Vol.~8., p.~161
\bibitem[\protect\citeauthoryear{M\"ucke \& Protheroe}{M\"ucke \& Protheroe}{2001}]{MueckeProtheroe}M\"ucke A., Protheroe R.~J., 2001, \aph, 15, 121
\bibitem[\protect\citeauthoryear{Nieppola, Tornikoski \& Valtaoja}{Nieppola et~al.}{2006}]{Nieppola}Nieppola E., Tornikoski M., Valtaoja E., 2006, \aap, 445, 441
\bibitem[\protect\citeauthoryear{Nikishov}{Nikishov}{1962}]{nikishov}Nikishov A.~I., 1962, Sov. Phys. JETP, 14, 393
\bibitem[\protect\citeauthoryear{{Nishijima}}{{Nishijima}}{2002}]{2002PASA...19...26N} {Nishijima} K. (CANGAROO collaboration), 2002, PASA, 19, 26
\bibitem[\protect\citeauthoryear{Nishiyama et~al.,}{Nishiyama et~al.}{1999}]{Nishiyama1999}Nishiyama T., et~al. (Utah Seven Telescope Array collaboration), 1999, in Proceedings of the 26th International Cosmic Ray Conference Vol.~3, Salt Lake City, p.~370
\bibitem[\protect\citeauthoryear{{Piner} \& {Edwards}}{{Piner} \& {Edwards}}{2004}]{2004ApJ...600..115P}{Piner} B.~G., {Edwards} P.~G., 2004, \apj, 600, 115
\bibitem[\protect\citeauthoryear{{Piner}, {Unwin}, {Wehrle}, {Edwards}, {Fey} \& {Kingham}}{{Piner} et~al.}{1999}]{1999ApJ...525..176P}{Piner} B.~G., {Unwin} S.~C., {Wehrle} A.~E., {Edwards} P.~G., {Fey} A.~L., {Kingham} K.~A., 1999, \apj, 525, 176
\bibitem[\protect\citeauthoryear{Piner, Pant \& Edwards}{Piner et~al.}{2007}]{PinerAK}Piner B.~G., Pant N., Edwards P.~G., 2007, \apj, in press, preprint (arXiv:0801.2749) 
\bibitem[\protect\citeauthoryear{{Primack}, {Bullock} \& {Somerville}}{{Primack} et~al.}{2005}]{primack421}{Primack} J.~R., {Bullock} J.~S., {Somerville} R.~S., 2005, in {Aharonian} F.~A., {V{\"o}lk} H.~J., {Horns} D., eds, High Energy Gamma-Ray Astronomy, AIP Conf. Ser. 745, pp 23--33
\bibitem[\protect\citeauthoryear{Punch}{Punch}{2007}]{hessmonitoring}Punch M. (H.E.S.S. collaboration), 2007, in {Proceedings of the 30th International Cosmic Ray Conference}, Merida, preprint (arXiv:0710.4057, p.~106)
\bibitem[\protect\citeauthoryear{Punch et~al.,}{Punch et~al.}{1992}]{Punch1992}Punch M., et~al. (Whipple collaboration), 1992, \nat, 358, 477
\bibitem[\protect\citeauthoryear{Quinn et~al.,}{Quinn et~al.}{1996}]{Quinn1996}Quinn J., et~al. (Whipple collaboration), 1996, \apj, 456, L83
\bibitem[\protect\citeauthoryear{Raue, Benbow, Costamante \& Horns}{Raue et~al.}{2007}]{0229}Raue M., Benbow W., Costamante L., Horns D. (H.E.S.S. collaboration), 2007, in {Proceedings of the 30th International Cosmic Ray Conference}, Merida, preprint (arXiv:0710.4057, p.~134)
\bibitem[\protect\citeauthoryear{{Rebillot}, {Badran}, {Blaylock} et~al.,}{{Rebillot} et~al.}{2006}]{2006ApJ...641..740R}{Rebillot} P.~F., {Badran} H.~M., {Blaylock} G., et~al., 2006, \apj, 641, 740
\bibitem[\protect\citeauthoryear{Rees}{Rees}{1978a}]{ReesM87}Rees M.~J., 1978a, MNRAS, 184, 61\textsc{p}
\bibitem[\protect\citeauthoryear{{Rees}}{{Rees}}{1978b}]{1978Natur.275..516R}{Rees} M.~J., 1978b, \nat, 275, 516
\bibitem[\protect\citeauthoryear{Richstone et~al.,}{Richstone et~al.}{1998}]{Richstone-98}Richstone D., et~al., 1998, \nat, 395, A14
\bibitem[\protect\citeauthoryear{{Sbarufatti}, Treves, Falomo, Heidt, Kotilainen \& Scarpa}{{Sbarufatti} et~al.}{2006}]{Sbar}{Sbarufatti} B., Treves A., Falomo R., Heidt J., Kotilainen J., Scarpa R., 2006, AJ, 132, 1
\bibitem[\protect\citeauthoryear{Schroedter et~al.,}{Schroedter et~al.}{2005}]{Schroedter}Schroedter M., et~al. (Whipple collaboration), 2005, \apj, 634, 947
\bibitem[\protect\citeauthoryear{{Sikora}, {Begelman} \& {Rees}}{{Sikora} et~al.}{1994}]{1994ApJ...421..153S}{Sikora} M., {Begelman} M.~C., {Rees} M.~J., 1994, \apj, 421, 153
\bibitem[\protect\citeauthoryear{{Spada}, {Ghisellini}, {Lazzati} \& {Celotti}}{{Spada} et~al.}{2001}]{Spada} {Spada} M., {Ghisellini} G., {Lazzati} D., {Celotti} A., 2001, \mnras, 325, 1559
\bibitem[\protect\citeauthoryear{Spergel et~al.,}{Spergel et~al.}{2007}]{WMAPpaperpublished} Spergel D.~N., et~al., 2007, \apjs, 170, 377
\bibitem[\protect\citeauthoryear{Stecker, de Jager \& Salamon}{Stecker et~al.}{1992}]{sjs92} Stecker F.~W., de Jager O.~C., Salamon M.~H., 1992, \apj, 390, 49
\bibitem[\protect\citeauthoryear{Stecker, Malkan \& Scully}{Stecker et~al.}{2006}]{steckerNEW} Stecker F.~W., Malkan M.~A., Scully S.~T., 2006, \apj, 648, 774
\bibitem[\protect\citeauthoryear{Stecker, Baring \& Summerlin}{Stecker et~al.}{2007}]{sbs07} Stecker F.~W., Baring, M.~G., Summerlin E.~J., 2007, \apj, 667, L29
\bibitem[\protect\citeauthoryear{Steele et~al.,}{Steele et~al.}{2007}]{veritasmonitoring} Steele D. (VERITAS collaboration), Carini M. T., Charlot P., Kurtanidze O., Lahteenmaki A., Montaruli T., Sadun A. C.,  Villata M., 2007, in {Proceedings of the 30th International Cosmic Ray Conference}, Merida, preprint (arXiv:0709.3869)
\bibitem[\protect\citeauthoryear{Superina, Benbow, Boutelier, Dubus \& Giebels}{Superina et~al.}{2007}]{0548icrc} Superina G., Benbow W., Boutelier T., Dubus G., Giebels B. (H.E.S.S. collaboration), 2007, in {Proceedings of the 30th International Cosmic Ray Conference}, Merida, preprint (arXiv:0710.4057, p.~138)
\bibitem[\protect\citeauthoryear{{Tavecchio}, {Maraschi} \& {Ghisellini}}{{Tavecchio} et~al.}{1998}]{Tavecchio1998} {Tavecchio} F., {Maraschi} L., {Ghisellini} G., 1998, \apj, 509, 608
\bibitem[\protect\citeauthoryear{Teshima et~al.,}{Teshima et~al.}{2007}]{279} Teshima M., et~al. (MAGIC collaboration), 2007, in {Proceedings of the 30th International Cosmic Ray Conference}, Merida, preprint (arXiv:0709.1475)
\bibitem[\protect\citeauthoryear{Tluczykont, Shayduk, Kalekin \& Bernardini}{Tluczykont et~al.}{2007}]{tlucmad} Tluczykont M., Shayduk M., Kalekin O., Bernardini E., 2007, J. Phys. Conf. Ser., 60, 318
\bibitem[\protect\citeauthoryear{{Tremaine} et~al.,}{{Tremaine} et~al.}{2002}]{2002ApJ...574..740T} {Tremaine} S., et~al., 2002, \apj, 574, 740
\bibitem[\protect\citeauthoryear{Treves, Falomo \& Uslenghi}{Treves et~al.}{2007}]{treves-07} Treves A., Falomo R., Uslenghi M., 2007, \aap, 473, L17
\bibitem[\protect\citeauthoryear{Uchiyama, Urry, Cheung, Jester, van Duyne, Coppi, Sambruna, Takahashi, Tavecchio \& Maraschi}{Uchiyama et~al.}{2006}]{uchiyama} Uchiyama Y., Urry C.~M., Cheung C.~C., Jester S., van Duyne J., Coppi P., Sambruna R.~M., Takahashi T., Tavecchio F., Maraschi L., 2006, \apj, 648, 910
\bibitem[\protect\citeauthoryear{Urry \& Padovani}{Urry \& Padovani}{1995}]{UrryPadovani} Urry C.~M., Padovani P., 1995, PASP, 107, 803
\bibitem[\protect\citeauthoryear{{Venters} \& {Pavlidou}}{{Venters} \& {Pavlidou}}{2007}]{2007ApJ...666..128V} {Venters} T.~M., {Pavlidou} V., 2007, \apj, 666, 128
\bibitem[\protect\citeauthoryear{{Wagner}, Dorner, Goebel, Hengstebeck, Kranich, Mazin, Nowak \& Tescaro}{{Wagner} et~al.}{2007}]{1553icrc}{Wagner} R.~M., Dorner D., Goebel F., Hengstebeck T., Kranich D., Mazin D., Tescaro D. (MAGIC collaboration), Nowak N., 2007, in {Proceedings of the 30th International Cosmic Ray Conference}, Merida, preprint (arXiv:0711.1586)
\bibitem[\protect\citeauthoryear{{Woo} \& {Urry}}{{Woo} \& {Urry}}{2002a}]{2002ApJ...579..530W}{Woo} J.-H., {Urry} C.~M., 2002a, \apj, 579, 530
\bibitem[\protect\citeauthoryear{{Woo} \& {Urry}}{{Woo} \& {Urry}}{2002b}]{2002ApJ...581L...5W}{Woo} J.-H., {Urry} C.~M., 2002b, \apj, 581, L5
\bibitem[\protect\citeauthoryear{{Woo}, {Urry}, {van der Marel}, {Lira} \& {Maza}}{{Woo} et~al.}{2005}]{2005ApJ...631..762W}{Woo} J.-H., {Urry} C.~M., {van der Marel} R.~P., {Lira} P., {Maza} J., 2005, \apj, 631, 762
\bibitem[\protect\citeauthoryear{{Wu}, {Liu} \& {Zhang}}{{Wu} et~al.}{2002}]{2002AA...389..742W} {Wu} X.-B., {Liu} F.~K., {Zhang} T.~Z., 2002, \aap, 389, 742

\end{thebibliography}

\end{document}